\documentclass[prd,floatfix,preprintnumbers,letterpaper]{revtex4}
\usepackage{amsmath,amssymb,graphicx}
\usepackage{hyperref}

\usepackage{xcolor}
\usepackage{graphicx}
\usepackage{amssymb}
\usepackage{multirow,bigdelim}
\usepackage{epstopdf}

\begin{document}
\title{Multiple measurements of quasars acting as standard probes: model independent calibration and exploring the Dark Energy Equation of States}
\author{Xiaogang Zheng$^{1}$, Shuo Cao $^{2\ast}$, Marek Biesiada$^{2,3}$, Xiaolei Li$^{4}$, Tonghua Liu
$^{2}$, and Yuting Liu$^{2}$}

\affiliation{1.School of Electrical and Electronic Engineering, Wuhan Polytechnic University, Wuhan 430023, China;\\
2. Department of Astronomy, Beijing Normal University,
100875, Beijing, China; \emph{caoshuo@bnu.edu.cn}\\
3. National Centre for Nuclear Research, Pasteura 7, 02-093 Warsaw, Poland;\\
4. College of Physics, Hebei Normal University, Shijiazhuang 050024, China}

\begin{abstract}
Recently, two classes of quasar samples were identified,
which are promising as new cosmological probes extending to higher
redshifts. The first sample uses the nonlinear relation between the
ultraviolet and X-ray luminosities of quasars to derive luminosity
distances, whereas the linear sizes of compact radio quasars in the
second sample can serve as standardized rulers, providing
angular-diameter distances. In this study, under the assumption of a
flat universe, we refreshed the calibration of multiple measurements
of high-redshift quasars (in the framework of a
cosmological-model-independent method with the newest Hubble
parameters data). Furthermore, we placed constraints on four models
that characterize the cosmic equation of state ($w$). The obtained
results show that: 1) the two quasar samples could provide promising
complementary probes at much higher redshifts, whereas compact radio
quasars perform better than ultraviolet and X-ray quasars at the
current observational level; 2) strong degeneracy between the cosmic
equation of state ($w$) and Hubble constant ($H_0$) is revealed,
which highlights the importance of independent determination of
$H_0$ from time-delay measurements of strongly lensed Quasars; 3)
together with other standard ruler probes, such as baryon acoustic
oscillation distance measurements, the combined QSO+BAO measurements
are consistent with the standard $\Lambda$CDM model at a constant
equation of state $w=-1$; 4) ranking the cosmological models, the
polynomial parametrization gives a rather good fit among the four
cosmic-equation-of-state models, whereas the
Jassal-Bagla-Padmanabhan (JBP) parametrization is substantially
penalized by the Akaike Information Criterion and Bayesian
Information Criterion criterion.
\end{abstract}

\maketitle

\section{Introduction}\label{section1}

Observing the type-Ia Supernovae has revealed the accelerated
expansion of the universe \cite{Riess1998, Perlmutter1999}, and the
follow-up projects have further supported the idea \cite{Suzuki2012,
Betoule2014, Scolnic2018}. This phenomenon is supported by the
increase in observational evidence such as the cosmic microwave
background (CMB) anisotropies \cite{WMAP,Planck2016, Planck2018},
baryon acoustic oscillation (BAO) \cite{Eisenstein2005, BOSS2017},
Hubble parameters derived from passively evolving galaxies
\cite{Moresco2016, Ratsimbazafy2017}, and Einstein radius
measurements of strong lensing systems \cite{Cao2012,Cao2015}. The
simplest way to describe the cosmic acceleration phenomenon is to
adopt the $\Lambda$CDM model, in which the cosmological constant
$\Lambda$ represents the negative-pressure dark energy (DE).
However, there are still the same problems under the assumption of
this simple model, including the well-known theoretical difficulties
of coincidence and the fine-tuning problems. From the observational
perspective, the values of the Hubble constant obtained from the
measurements of local supernova through the SH0ES collaboration
\cite{Riess2016,Riess2019} are in tension with the {\it Planck} CMB
data analyzed under the assumption of a flat $\Lambda$CDM model
\cite{Planck2016,Planck2018}. The $\rm H_0$ value derived from
gravitational lensing time-delay distances \cite{Wong2020} under the
assumption of a flat $\Lambda$CDM model is also in tension with the
estimates of the early universe \cite{Planck2018}. Recently, some
other observational data, including the Hubble parameter
measurements \cite{Zheng2016} and quasars calibrated as standard
candles \cite{Risaliti2019}, were observed to be in tension with the
$\Lambda$CDM model, especially in the so-called "redshift desert"
region. Moreover, another important issue is whether the DE
equation-of-state (EoS) parameter evolves with redshifts
\cite{Cao2014}.

Currently, most of the cosmological observations refer to either low
\cite{Suzuki2012, Betoule2014, Scolnic2018} or high redshift
\cite{WMAP,Planck2016, Planck2018} regions. The so-called "redshift
desert" makes it difficult to study the dynamical evolution of DE.
Consequently, a promising probe could be quasars, which are the
brightest objects in the universe seen at very high redshifts.
Although using quasars as standard cosmological probes is ambiguous,
it was proposed that luminosity distances can be assessed from the
radius of the broad-line region (BLR)-the monochromatic luminosity
relation \cite{Watson2011,Aldama2019}, the properties of the
super-Eddington accreting massive black holes \cite{Wang2013}, the
correlation between the X-ray variability amplitude and the mass of
the black hole \cite{LaFranca2014}, or the nonlinear relation
between the ultraviolet (UV) and X-ray luminosities
\cite{Risaliti2015,Risaliti2019}. Meanwhile, the angular-diameter
distances can be derived from the classical geometrical size of the
BLR \cite{Elivs2002}, the angular size-redshift relation of compact
structures in intermediate-luminosity radio quasars
\cite{Cao2017a,Cao2017b,Cao2018}, etc. Taking advantage of the
available rich samples, we employed the nonlinear relation between
the UV and X-ray flux measurements of quasars \cite{Risaliti2019},
and the angular size-redshift relation from compact radio quasars
\cite{Cao2017a}. Besides the large dispersion of the data points,
these two compilations of quasar suffer from the ambiguity in the
meaning of nuisance parameters, which should be calibrated with
external probes. However, benefiting from their broad redshift
coverage, they have the potential to address the issues of the
dynamical evolution of DE, the measurements of the speed of light
\cite{Cao2020}, and cosmic curvature at higher redshifts
\cite{Cao2019}, and the testing of the cosmic-distance-duality
relation in the early universe \cite{Zheng2020}.

In this study, we take advantage of the high redshift coverage for
both the Hubble parameters derived from differential ages of
passively evolving galaxies (so-called cosmic chronometers)
\cite{Moresco2016, Ratsimbazafy2017} and the two different kinds of
quasar samples\cite{Risaliti2019,Cao2017a}. First, we refreshed the
calibration of the nuisance parameters in the nonlinear relation
between the UV and X-ray luminosity of quasars (QSO[XUV] hereafter)
and that in the angular size-redshift relation of the compact
structure sizes in radio quasars (QSO[CRS]). This was performed in a
nonparametric, model-independent technique developed by
ref.\cite{Seikel2012}. Because the redshift coverage of the Hubble
parameters ($0.07<z<1.965$) is smaller in both quasar samples,
$0.462<z<2.73$ for QSO[CRS] \cite{Cao2017a} and $0.036<z<5.1003$ for
QSO[XUV] \cite{Risaliti2019}, we extended the calibrated nuisance
parameters and applied them to the redshift range larger than the
$H(z)$ sample. We compared two quasar samples with regard to their
ability to constrain cosmological model parameters. In particular,
we studied different parametrizations of the cosmic equation of
state (EoS), combined with the current BAO data and Hubble constant
priors.

In Section \ref{sec:calibration}, we introduce two kinds of quasar
samples and calibrate their nuisance parameters. Then, in Section
\ref{sec:constraint}, we introduce four parametrizations of DE EoS
allowed to evolve in redshifts and test them with the calibrated
quasar samples combined with the BAO data. The results are discussed
in Section \ref{sec:discussion} and the conclusions are presented in
Section \ref{sec:conclusion}.

\section{Model Independent Calibration of Quasar Samples}\label{sec:calibration}

The redshift range probed by quasar measurements has a wide
coverage, but such data can hardly be used directly as a standard
cosmological probe because quasars have neither standard luminosity
nor a clearly defined size scale. Recently, interesting methods that
provide the luminosity distance and angular-diameter distance by
quasars through multiband observations have been proposed, including
the nonlinear relation between the UV and X-ray luminosities of
quasars \cite{Risaliti2015,Risaliti2019} and the angular
size-redshift relation of the compact radio structure
\cite{Cao2017a,Cao2017b}, respectively.

According to the empirical relation between the UV and X-ray
luminosities of quasars, $\log(L_X)=\gamma \log(L_{UV})+\beta$ (for
details, see ref.\cite{Risaliti2015,Risaliti2019}), and the
flux-luminosity relation of $F_{\nu}=L_{\nu}/4\pi D_L^2$, one can
relate the respective fluxes (observable quantities) with the
luminosity distance thus:
\begin{align}\label{Eq:logFx}
\log(F_X) & \equiv \Phi(\log F_{UV},D_L) \nonumber\\
 & =\beta'+\gamma \log(F_{UV})+2(\gamma-1)\log(D_L(z;\textbf{p}))
\end{align}
where $\rm\beta'=\beta+(\gamma-1)log(4\pi)$ is the effective
intercept, which is defined as the combination of the intercept
$\beta$ and the slope $\gamma$. In this formula, the unit of flux
densities $\rm F_X$ and $\rm F_{UV}$ is $\rm
erg\,s^{-1}\,cm^{-2}\,Hz^{-1}$, and that luminosity distances is
$\rm Mpc$. In principle, the luminosity distance in
Eq.(\ref{Eq:logFx}) is the true one and can be confronted with the
theoretically predicted one $D_L(z;\textbf{p})$, where $\textbf{p}$
denotes all cosmological model parameters. To test cosmological
models, one needs the prior knowledge of the $\rm\beta$ and $\gamma$
coefficients. In other words, one needs to calibrate this relation
in a cosmological model independent way. As discussed in a later
section, we employed $D_L(z)$ obtained from $H(z)$ measurements
provided by cosmic chronometers. Another option could be a joined
assessment of the nuisance parameters with the cosmological
parameters during the
investigation\cite{Melia2019,Zheng2020,Khadka2020a,Khadka2020b,Liu2020,Wei2020}.
As discussed in \cite{Risaliti2015}, a slope parameter $\gamma$ can
be fitted on the subsamples covering different redshift bins across
the full redshift range of the data. It has the advantage of being
able to test whether the $\gamma$ parameter evolves with redshifts,
and if it does not, one can estimate the effective slope as an
average over the bins. It is different in the case of the intercept
parameter $\beta$, which is hard to check without having a deep
physical understanding of the nonlinear luminosity relation.
Moreover, the data points are still largely scattered despite the
that, in the compiled sample used herein, the scatter has already
been significantly reduced. Therefore, in the statistical analysis,
we also included another nuisance parameter $\delta$, which
describes this scatter.

To obtain reliable information on the cosmological model, a
reasonable prior of nuisance parameters calibrated from more precise
external probes is necessary. We optimized the nuisance parameters
using the luminosity distances from the model independent of the
Gaussian Process by minimizing the likelihood the function
\begin{align}\label{eq:Lik_XUV}
\ln \mathcal{L}_{QSO[XUV]}= & -\frac{1}{2}\sum^N_i\nonumber\\
& \left\lbrace\frac{[\log(F_X)_i-\Phi(\log
F_{UV},D_L)]^2}{s^2_i}+\ln(2\pi s_i^2)\right\rbrace
\end{align}
where the expression $\Phi(F_{UV},D_L)$ is given in
Eq.(\ref{Eq:logFx}) and the variance, $s_i^2=\sigma_i^2+\delta_i^2$,
is the combination of $\sigma_i$ and $\delta_i$, which denote the
statistical uncertainties and the intrinsic scatter, respectively.
To calibrate the nuisance parameters, ($\gamma$ and $\beta$), the
statistical uncertainties include $\sigma(\log F_X)$ and
$\sigma(D_L)$, which denote the uncertainties of the observed $\log
F_X$ and reconstructed $D_L$, respectively. Namely, $\sigma_i^2=
\sigma(\log F_X)_i^2+(\frac{2(\gamma-1)}{D_L\ln10}\sigma(D_L))_i^2$.
For the investigation of the DE EoS, the statistical uncertainty
include $\sigma(\log F_X)$ and the gaussian priors are assumed for
the nuisance parameters. As discussed in ref.\cite{Risaliti2019},
the uncertainty of $F_{UV}$ is negligible compared to that of $F_X$,
and consequently, it is ignored herein. In ref.\cite{Risaliti2015},
the "best sample" of 808 quasars was extracted for cosmology
inference, and the sample was enlarged to 1598 in
ref.\cite{Risaliti2019}, and that was employed in this study.

On the other hand, another quasar sample obtained through radio
observations was proposed to provide angular-diameter distances
using the simple angular size ($\theta$)-redshift ($z$) relation
\cite{Kellermann1993,Gurvits1994,Gurvits1999}. The angular size at
different redshifts $\theta(z)$ can be expressed in terms of the
intrinsic metric size $l_m$ of the compact radio structure of
quasars and the angular-diameter distances $D_A(z)$ as follows:
\begin{equation}\label{Eq:thetaz}
\theta(z)=\frac{l_m}{D_A(z)}
\end{equation}
where $l_m$ may depend on the luminosity of the source or evolve
with redshifts. In ref.\cite{Cao2017a,Cao2017b}, a sample of 120
intermediate-luminosity radio quasars with negligible dependence on
both luminosity and redshift was identified. Thus, the intrinsic
metric length $l_m$ can be taken simply as the linear size parameter
$l$. To calibrate the linear size using the cosmological model
independent $H(z)$ data, we maximized the likelihood function
defined as:
\begin{equation}\label{eq:Lik_CRS}
\ln\mathcal{L}_{QSO[CRS]}=-\frac{1}{2}\sum^{120}_{i=1}\frac{[\theta_{obs}(z_{i})-\theta_{th}(z_{i}]^{2}}{\sigma_{\theta(z_i)}^{2}}
\end{equation}
where the variance,
$\sigma_{\theta(z_{i})}^2=[\sigma^{sta}_{\theta(z_{i})}]^2+[\sigma^{sys}_{\theta(z_{i})}]^2$,
includes the statistical and $10\%$ systematic uncertainties. This
method is abbreviated as the subscript, QSO[CRS]. To calibrate the
linear-size parameter, the statistical uncertainties include two
terms: one from the observed angular size and the other propagated
from the reconstructed angular-diameter distance, i.e.,
$[\sigma_{\theta(z_{i})}]^2=[\sigma(\theta_{obs}(z_i))]^2+[\frac{l}{D_A^2(z_i)}
\sigma(D_A(z_i))]^2$. To investigate DE EoS, rather than the
uncertainty of the reconstructed angular-diameter distance
$\sigma(D_A)$, that of the calibrated linear size $\sigma(l)$ was
considered. Such methodology has been extensively employed in
subsequent cosmological studies of intermediate-luminosity radio
quasars \cite{Qi2017,Zheng2017,Xu2018}

Lastly, as already mentioned above, the cosmic expansion rates
$H(z)$ were used to calibrate the nuisance parameters. The sample
used was obtained through the cosmological-model-independent method
using the so-called cosmic chronometers \cite{Moresco2016,
Ratsimbazafy2017}. The idea is that
$H(z)=\frac{\dot{a}}{a}=-\frac{1}{1+z}\frac{dz}{dt}$, which is
expressed in terms of observable quantities (i.e., redshifts and
their change over time), could be obtained without assuming any
particular cosmological model. The point here is to measure $dz/dt$
directly, which could be achieved using massive, early-type galaxies
that evolve passively on a timescale longer than their age
difference. Certain features of their spectra, such as D4000 that
breaks at 4000{\AA}, which indicates the evolution of their stellar
populations, enable us to measure the age difference of such
galaxies. As discussed in ref.\cite{Cao2017a}, the choice of a
stellar-population-synthesis model may strongly affect the estimates
of the age difference at higher redshifts $z>1.2$. Hence, only 24
data points, up to $z<1.2$, were considered in ref.\cite{Cao2017a}.
The redshift coverage of both the above-mentioned quasar samples is
much broader than $H(z)$ measurements. Moreover, such a restriction
would exclude a considerable fraction of angular size $\theta(z)$
data and could affect the results of the calibrations. Therefore, we
exploited a wider redshift range and used a total of 31 $H(z)$
measurements with redshifts of $0.07<z<1.965$.

\begin{figure}
\begin{center}
\centering
\includegraphics[angle=0,width=85mm]{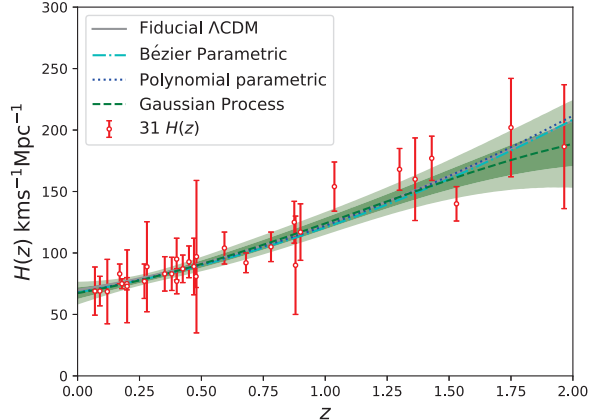}
\caption{\label{fig:GP_Hz} The observed Hubble parameters and
reconstructed expansion rate up to the redshift $z=2$. The red
points with error bars represent the $H(z)$ measurements obtained
using cosmic chronometers and the green dashed line indicates the
$H(z)$ function reconstructed with Gaussian Processes, and it is
surrounded by the green region of various transparencies
corresponding to the $1\sigma$ and $2\sigma$ confidence region. For
comparison, we added the theoretical prediction from a fiducial flat
$\Lambda$CDM model with the Hubble constant $\rm
H_0=70\;km\,s^{-1}\,Mpc^{-1}$ and matter-density parameter
$\Omega_m=0.3$ (gray line), the reconstructed $H(z)$ from the
B\'ezier parametric curve (cyan dash-dot line), and that from
log-polynomial parameterization (blue dotted line).}
\end{center}
\end{figure}

From the observed data, we reconstructed the $H(z)$ function in the
redshift range $0<z<2$ using a model-independent method of Gaussian
Processes (GP). The Python package, GaPP, developed in
ref.\cite{Seikel2012} and used in ref.\cite{Yin2019,Zheng2020}, was employed in the reconstruction, assuming
zero-mean and squared exponential-covariance functions. The effects
of the mean function and covariance selection were discussed in our
previous study \cite{Zheng2020}, where it was shown that the choice
of the covariance function does not have a significant influence on
the results, and the impact of the mean-function choice was even
smaller. The results are shown in Fig.(\ref{fig:GP_Hz}), where the
red points with the error bars, the green dashed line, and the green
regions with different transparency indicates the 31 $H(z)$
measurements, GP reconstructed $H(z)$, and corresponding confidence
regions, respectively. Moreover, different reconstruction methods
may also affect the results and subsequent calibration. To show the
impact of such methods, we performed reconstruction separately using
the B\'ezier parametric curve of degree $n=2$, which was used to
investigate the Amati relation \cite{Amati2019}, and the
log-polynomial form employed in QSO[XUV] studies
\cite{Risaliti2019}. More specifically,
$H^{Bez}(z)=\beta_0(1-\frac{z}{z_m})^2+2\beta_1\frac{z}{z_m}(1-\frac{z}{z_m})+\beta_2(\frac{z}{z_m})^2$
and $H^{Pol}(z)=H_0[\log(1+z)+a_2\log^2(1+z)+a_3\log^3(1+z)]$. The
reconstruction results are shown in Fig.(\ref{fig:GP_Hz}), where the
cyan dash-dot line indicates the $H(z)$ reconstructed from the
B\'ezier parametric curve and the blue dotted line indicates that
from the log-polynomial parametrization. Their confidence regions
are not shown so as not to blur the picture since they are similar
to that from the Gaussian process. The theoretical prediction from a
fiducial flat $\Lambda$CDM model with the Hubble constant $\rm
H_0=70\;km\,s^{-1}\,Mpc^{-1}$ and matter-density parameter $\rm
\Omega_m=0.3$ was also performed for comparison. Despite that the
reconstructed $H(z)$ was more consistent with the theoretical
prediction from the fiducial model, both of them encountered the
problem of order truncation, which may affect the results,
especially at high redshifts \cite{Banerjee2020,Yang2020}, and we
preferred to use the GP method in this study. Through GP, we also
obtained the Hubble constant $\rm H(z=0)=67.32\pm4.7\,km\;s^{-1}\;Mpc^{-1}$, which will be used later.
The result of the reconstruction is consistent with that using only
the 31 cosmic chronometric measurements reported in \cite{Yu2018,
Liu2020}. Furthermore, combining these 31 $H(z)$ measurements with
the latest SN Ia compilations (Pantheon and MCT), ref.\cite{Valent2018}
obtained a more accurate value for the Hubble constant with the best
value consistent with that of ref.\cite{Yu2018, Liu2020}, whereas we
only considered the application of 31 $H(z)$ measurements herein.

Once we obtain the reconstructed $H(z)$, we can derive the
cosmological distances. As reported in ref.\cite{Liao2013}, under
the assumption of flat universe, the comoving distance can be
calculated from the reconstructed $H(z)$ using the usual simple
trapezoidal rule:
\begin{equation}\label{eq:Hz_D}
D(z)=c\int^z_0\frac{dz'}{H(z')}\approx\frac{c}{2}\sum^n_{i=1}(z_{i+1}-z_i)\left[\frac{1}{H(z_{i+1})}+\frac{1}{H(z_i)}\right]
\end{equation}
and the corresponding uncertainty is given by
\begin{equation}\label{eq:Hz_dD}
\delta
D(z)=\frac{c}{2}\sum^n_{i=1}(z_{i+1}-z_{i})\left(\frac{\sigma^2_{H_{i+1}}}{H^4_{i+1}}+\frac{\sigma^2_{H_{i}}}{H^4_{i}}\right)^{1/2}
\end{equation}
Then, the angular-diameter and luminosity distances can be
calculated as $D_A(z)=D(z)/(1+z)$ and $D_L(z)=D(z)(1+z)$,
respectively.

\begin{figure*}
\begin{center}
\centering
\includegraphics[angle=0,width=90mm]{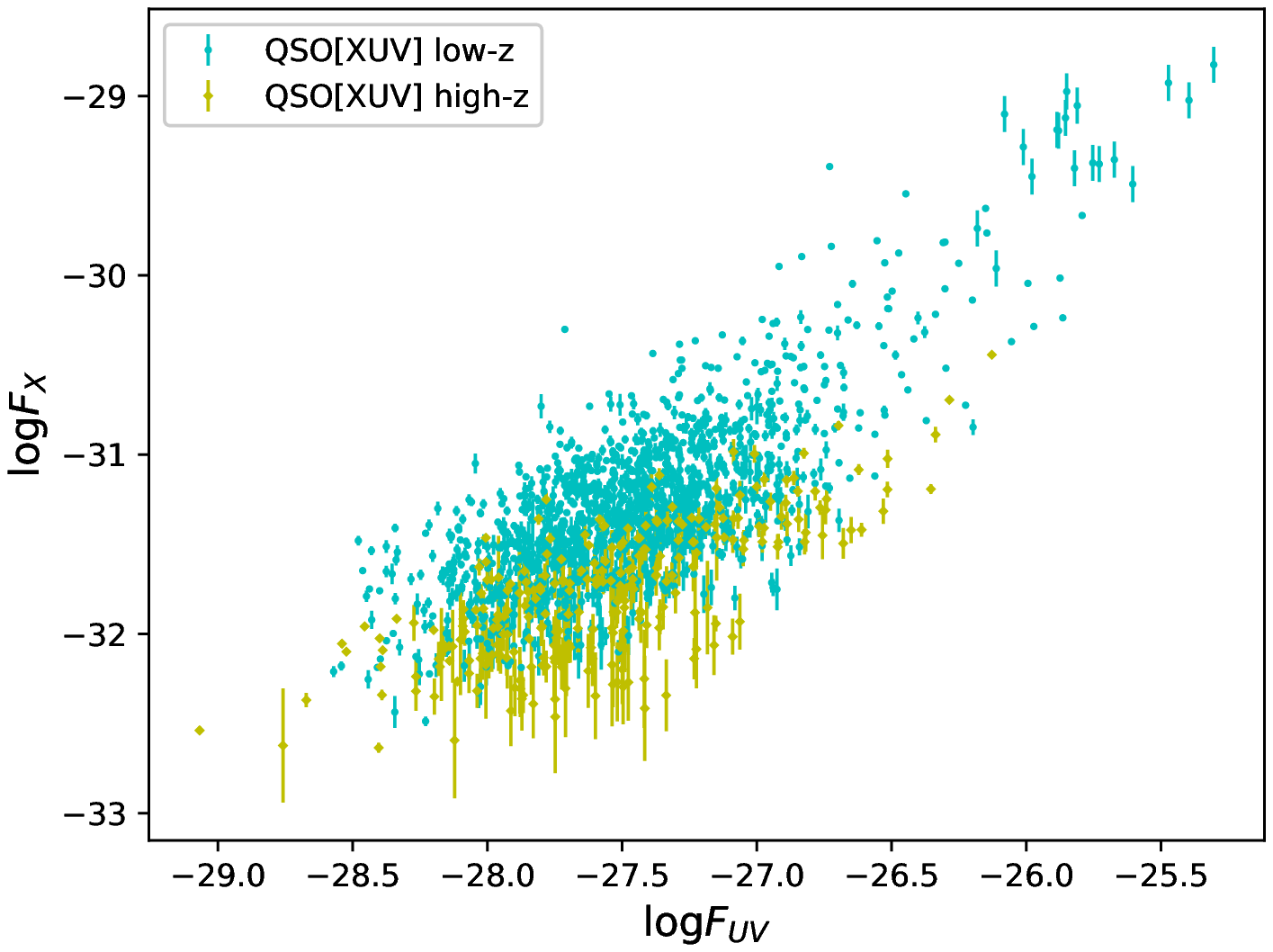}
\includegraphics[angle=0,width=65mm]{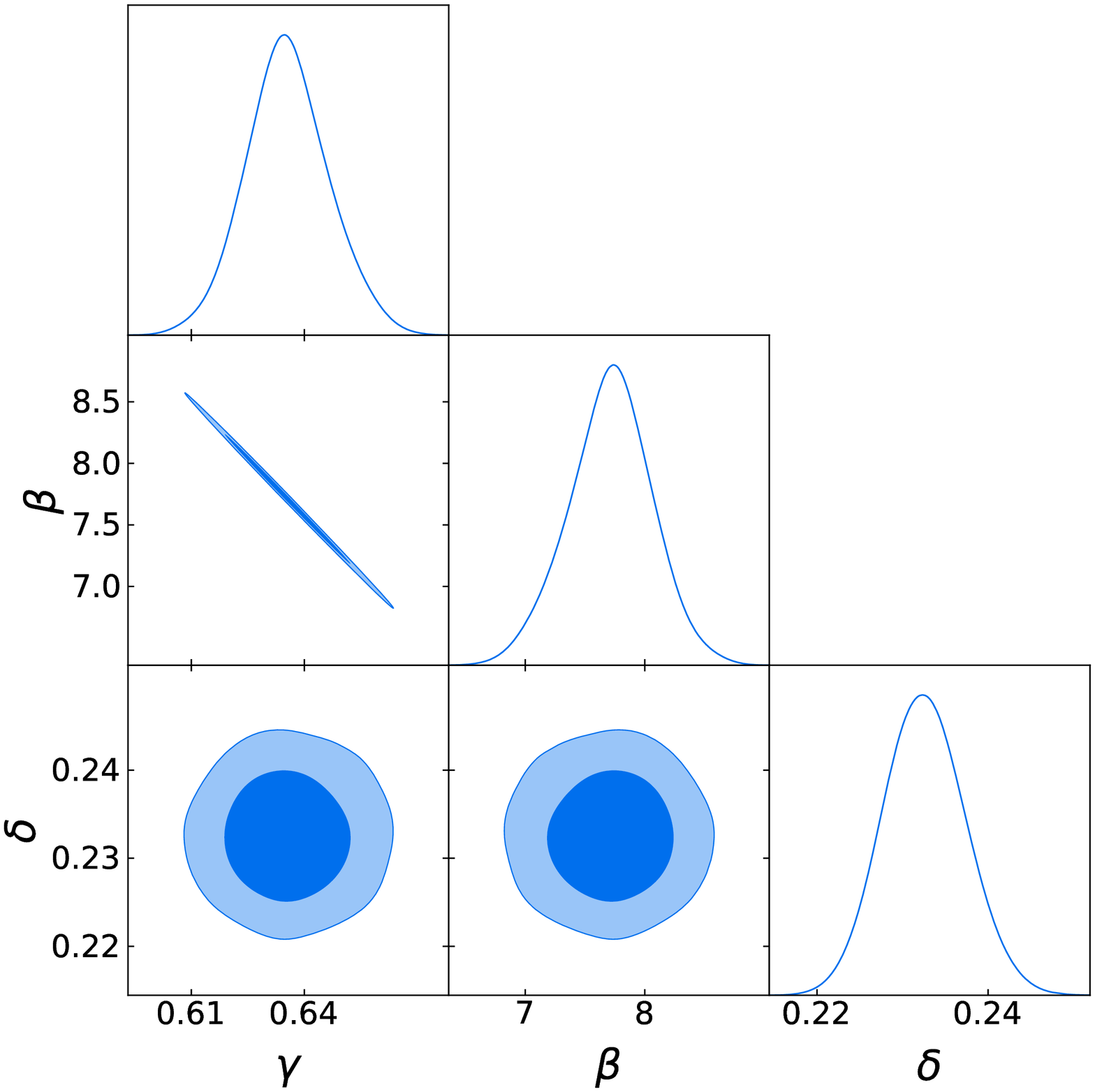}
\caption{\label{fig:cal_XUV} Calibration of QSO[XUV] nuisance
parameters with external $H(z)$ measurements. Left: The flux
relation between $\log F_X$ and $\log F_{UV}$, where the cyan circle
points correspond to the data matched with the reconstructed
$H(z<2.0)$ used during the calibration. Yellow diamond points
represent the rest of the sample. Right: Two-dimensional confidence
regions and marginal distributions for the calibrated parameters:
the slope parameter $\gamma=0.634\pm0.011$, the intercept
$\beta=7.75\pm0.34$ and dispersion $\delta=0.232\pm0.005$ for
QSO[XUV] sample.}
\end{center}
\end{figure*}

\begin{figure*}
\begin{center}
\centering
\includegraphics[angle=0,width=90mm]{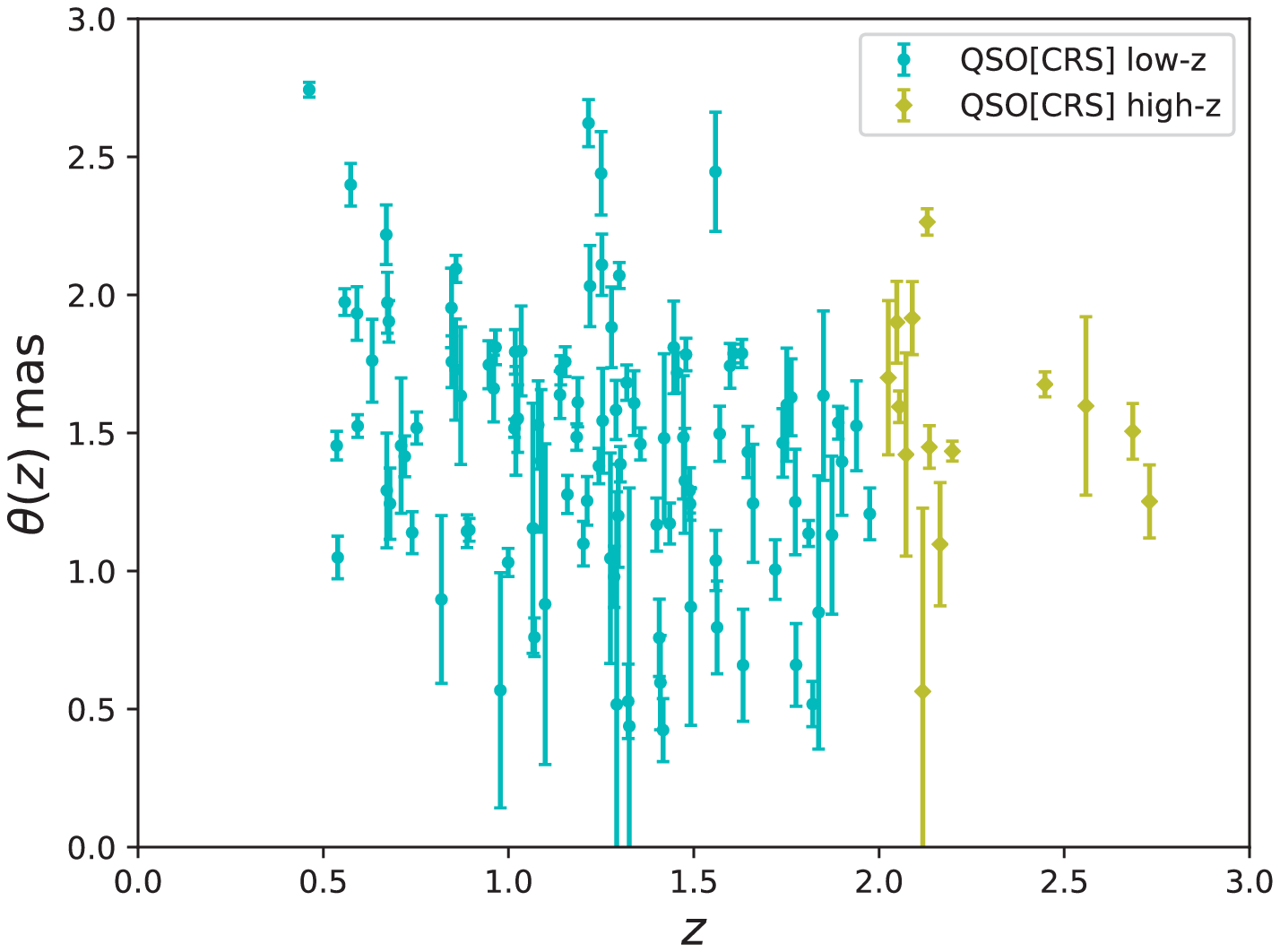}
\includegraphics[angle=0,width=60mm]{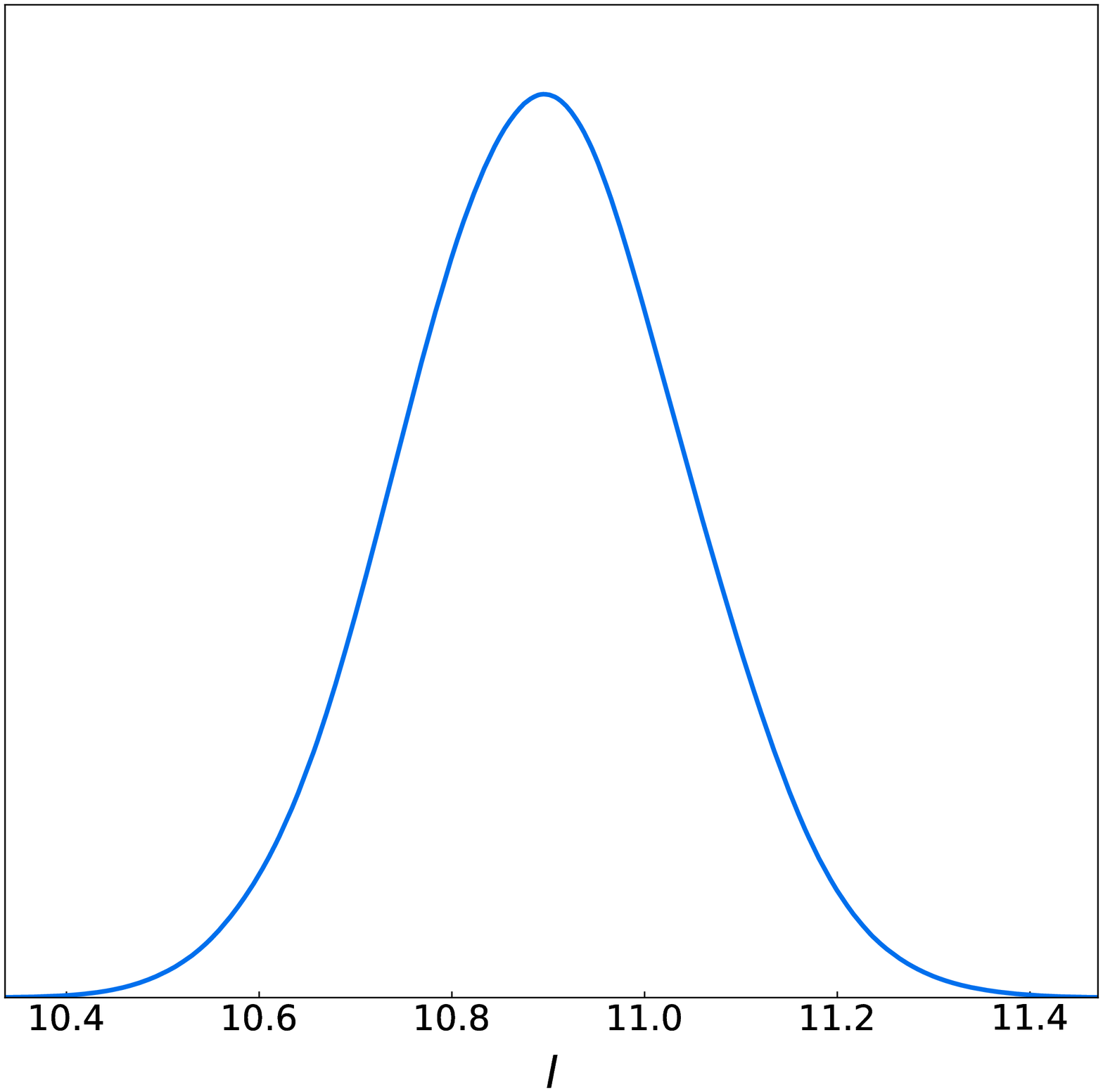}
\caption{\label{fig:cal_CRS} Calibration of QSO[CRS] linear size
parameter with external $H(z)$ measurements. Left: Observed angular
size-redshift relation of intermediate-luminosity quasars, where the
points matched with the reconstructed $H(z<2.0)$ are displayed in
cyan circle. Yellow diamond points represent the rest of the sample.
Right: Probability distribution of the best-fitted linear size
parameter $l=10.89\pm0.19$ pc.}
\end{center}
\end{figure*}

\begin{figure*}
\begin{center}
\centering
\includegraphics[angle=0,width=80mm]{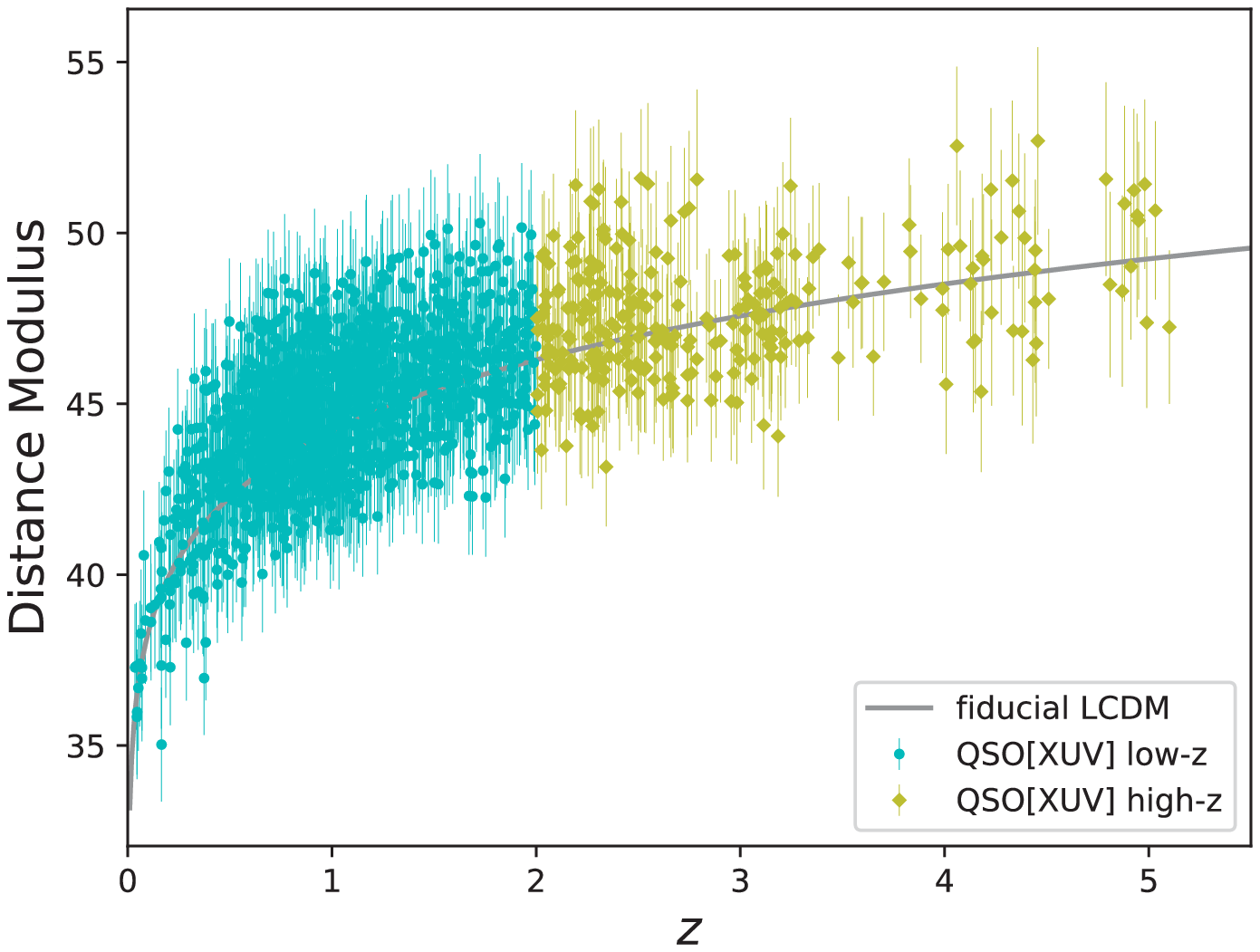}
\includegraphics[angle=0,width=80mm]{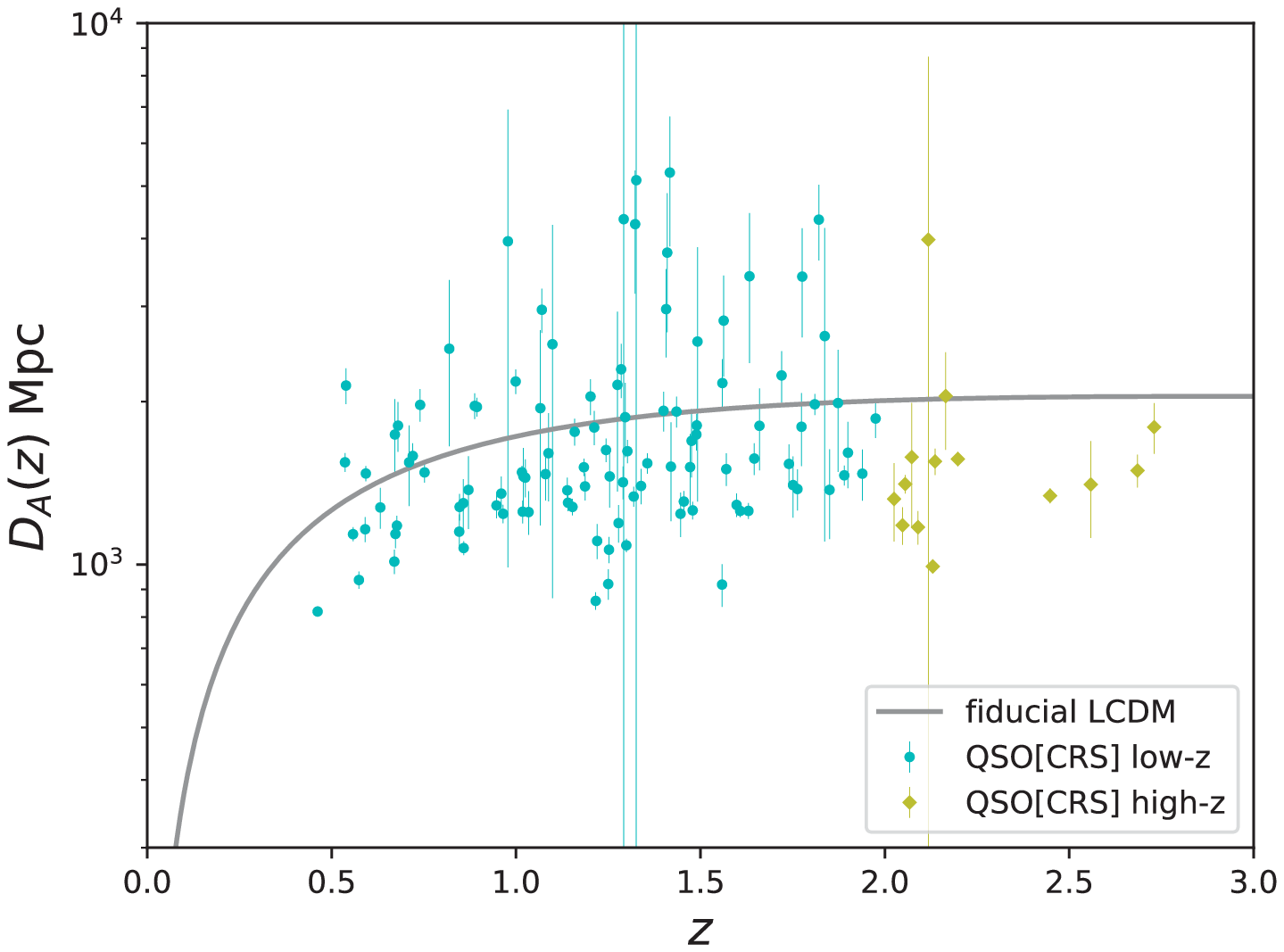}
\caption{\label{fig:cal_distance} Distance modulus of QSO[XUV]
(left) and angular-diameter distances of QSO[CRS] (right) calculated
using the prior values of the calibrated nuisance parameters. The
cyan circle points correspond to the data used for the calibration
of the nuisance parameter and the yellow diamond points represent
the higher-redshift sample. The theoretical prediction from the
fiducial flat $\Lambda$CDM model with the Hubble constant $\rm
H_0=70\;km\,s^{-1}\,Mpc^{-1}$ and density parameter $\rm
\Omega_m=0.3$ is indicated by the gray line for comparison.}
\end{center}
\end{figure*}

Matching the distances reconstructed from $H(z)$ with the data from
both QSO samples, the totals of 1330 and 106 data pairs were
selected for QSO[XUV] and QSO[CRS], respectively. The samples were
calibrated with a maximum likelihood approach based on
Eq.(\ref{eq:Lik_XUV}) and Eq.(\ref{eq:Lik_CRS}), as illustrated in
Fig.(\ref{fig:cal_XUV}) and Fig.(\ref{fig:cal_CRS}). The best-fitted
values of the slope parameter $\gamma=0.634\pm0.011$, intercept
parameter $\beta=7.75\pm0.34$, and dispersion $\delta=0.232\pm0.005$
for the QSO[XUV] sample were considered as the calibrated results
and employed in the investigation of DE EoS as the Gaussian priors.
Comparing with some previous works, the results obtained in study
are consistent with the $\gamma=0.633\pm0.002$ and $\delta=0.24$
from a best-fit log-linear model in ref.\cite{Risaliti2019} and the
$\gamma=0.62\pm0.01$ and $\delta=0.23\pm0.004$ obtained under the
assumption of a flat $\Lambda$CDM model by ref.\cite{Khadka2020b}.
The intercept parameter $\beta$ is fully dependent on external
calibrators; hence, it is harder to compare. However, the obtained
value is consistent with that of the fits reported in
ref.\cite{Khadka2020b}. For direct cosmological investigations, one
needs to determine the cosmological distances at the corresponding
redshifts. For the 1598 QSO[XUV] sample \cite{Risaliti2019} used in
this study, the luminosity distance can be computed using
Eq.(\ref{Eq:logFx}) as
\begin{equation}\label{Eq:logDL}
\log D_L=\frac{[\log F_X-\gamma\log
F_{UV}-\beta-(\gamma-1)\log(4\pi)]}{2(\gamma-1)}
\end{equation}
 The distance modulus is given by
\begin{align}\label{Eq:mu}
\mu = & 5\log D_L+25 \nonumber\\
    = & \frac{5}{2(\gamma-1)}[\log F_X-\gamma\log F_{UV}-\beta-(\gamma-1)\log(4\pi)]+25
\end{align}
and the corresponding uncertainty is given by
\begin{eqnarray}\label{Eq:dmu}
\sigma(\mu)= & \frac{5}{2(\gamma-1)}\bigg[\sigma(\log F_X)^2+ \sigma(\beta)^2 \nonumber\\
      +&\left(\frac{\log F_{UV}-\log F_X+\beta}{\gamma-1}\sigma(\gamma) \right)^2 \bigg]^{1/2}
\end{eqnarray}
The derived distance modulus and the corresponding uncertainties are
shown in the left panel of Fig.(\ref{fig:cal_distance}), where
$\sigma(\gamma)$ and $\sigma(\beta)$ are taken as $1\sigma$
uncertainties from the aforementioned calibration results, as shown
in the right panel of Fig.(\ref{fig:cal_XUV}). Owing to the large
dispersion of distance modulus, another scatter parameter could be
introduced for the likelihood calculation of $\mu$. In this study,
we preferred to use the likelihood Eq.(\ref{eq:Lik_XUV}) directly
with the Gaussian priors on the nuisance parameters ($\gamma$ and
$\beta$) and intrinsic scatter ($\delta$).

For the linear size, the best-fitted value was $l=10.89\pm0.19\,pc$.
It is somewhat smaller than but marginally consistent with that
calibrated with the Union2.1 SN Ia compilation,
$l=11.42\pm0.28\,pc$, in ref.\cite{Cao2017a} and that calibrated on
a subsample of $H(z)$ measurements covering redshifts $z<1.2$ in
ref.\cite{Cao2017b} ($l=11.03\pm0.25$). With the measurements of the
angular size and the calibrated linear size, the angular-diameter
distance and the corresponding uncertainties can be calculated as
$D_A(z)=l/\theta(z)$ and $\sigma(D_A) =
\left[(\sigma(l)/\theta)^2+(l \sigma(\theta)/\theta^2)^2
\right]^{1/2}$. The calculated results are shown in the right panel
of Fig.(\ref{fig:cal_distance}).

Comparing the measurements of the lower ($0<z<2$) and the higher
redshifts ($z>2$), the $\log F_X$-$\log F_{UV}$ correlation of
QSO[XUV] appears redshift dependent (the left panel of
Fig.(\ref{fig:cal_XUV})) and the angular size $\theta(z)$ seems
larger and the angular-diameter distance $D_A(z)$ smaller than the
average value (the left panel of Fig.(\ref{fig:cal_CRS}) and the
right panel of Fig.~(\ref{fig:cal_distance}), respectively).
Considering the uncertainties of $\log F_X$, the slopes are similar,
and the intercepts are different. This is consistent with the
evolution discussion of the nuisance parameters in
ref.\cite{Risaliti2019} (their supplementary Fig.~8). The lack of
$\theta(z)$ measurements at higher redshifts may affect the QSO[CRS]
approach. It should be noted that the sparsity of the current $H(z)$
data at higher redshifts may affect the reliability of calibration,
particularly in light of using them outside the redshift range of
the actual calibration. Hence, there is hope that the data on Hubble
parameters would be extended in the future to higher redshift
ranges.

\section{Investigation of the Dark Energy Equation of State}\label{sec:constraint}

DE is synonymous with the unknown cause of the accelerated expansion
of the universe. Since it is still unknown and we strive toward
resolving the enigma, a useful approach is to model it as a fluid
with some EoS: $p=w \rho$, where $p$ is the pressure and $\rho$ the
energy density. The simplest model used in this context is the
$\Lambda$CDM model, where the cosmological constant $\Lambda$ -- an
entity having a long history in cosmology -- represents DE.
Formally, this was equivalent to $w=-1$. A somewhat more complicated
model of DE is the so-called $w$CDM model, where the $w$ coefficient
is also a constant but different from $\Lambda$. These models can be
classified into quintessence models ($-1<w<-1/3$), including phantom
models ($w<-1$) and quintom model, with $w$ crossing the $-1$ value.
Most of the observations are very well consistent with the
$\Lambda$CDM model, but significant deviation from it is still
allowed according to their accuracy and precision. The crucial issue
here is whether the cosmic EoS evolves in time (i.e., in redshift).
Such unambiguous finding would be of utmost importance for
theoretical studies, where a more fundamental explanation of DE is
sought for, and the only (or one of the very few) common points with
the real observational data could be the effective $w$ parameter,
not necessarily a constant. Below, we present four particular
parametrizations of evolving EoS in flat universe to be studied in
order to find out if and to what extent they are supported by QSO
data combined with BAO measurements.

\subsection{DE EoS parametrizations}

To investigate the dynamical evolution of DE, which could be
manifested as an EoS coefficient dependent on redshift $w(z)$,
several phenomenological parameterizations have been proposed. Since
none of them is clearly supported by a theoretical model, we chose
just four of them. The simplest phenomenology could be just a
first-order Taylor expansion: $w(z)=w_0+w_a z$. It is divergent at
redshifts $z>1$ and refers to the observable quantity $z$, which is
not a physical degree of freedom like the scale factor $a(t)$. The
same idea of Taylor expansion, but in $a(t)$ rephrased in terms of
$z$, led to the most widely used form of DE EoS, i.e., the
Chevallier-Polarski-Linder (CPL) parametrization
\cite{Chevallier2001,Linder2003}
\begin{equation}
w(z)=w_0+w_a\frac{z}{1+z}.
\end{equation}
This parametrization has several advantages \cite{Linder2003} such
as well-bounded behavior at high redshifts, simple physical
interpretations, and manageable two-dimensional phase space.
However, it is also associated with some problems: it will encounter
divergency in the future as $z\sim-1$; the various limit values at
low-z ($w(z\rightarrow0)\sim w_0$) and
high-z($w(z\rightarrow\infty)\sim w_0+w_a$), etc. Similar to the CPL
parametrization, the JBP parametrization \cite{Jassal2005} was
proposed as
\begin{equation}
w(z)=w_0+w_a\frac{z}{(1+z)^2}
\end{equation}
and it has the same DE EoS at present ($z\rightarrow 0$) or remote
past($z\rightarrow\infty$), with rapid variation at low-z. For
comparison, we also discuss the logarithmic parametrization
\cite{Efstathiou1999}\begin{equation} w(z)=w_0+w_a\ln(1+z)
\end{equation}
This becomes infinite when $z\rightarrow\infty$ under the assumption
of the DE EoS. So, it can describe the behavior of DE when the
redshift is not so high. Similarly to the method of fitting the
Hubble diagram with a polynomial function applied in
ref.\cite{Risaliti2019}, we investigated the polynomial
parametrization \cite{Weller2002}:
\begin{equation}
w(z)=-1+w_0(1+z)+w_a(1+z)^2
\end{equation}

The dimensionless expansion rate $E(z)$ can be calculated from the
Friedman equation. A combined expression for this (squared) in the
above described EoS parametrizations is as follows:
\begin{align}\label{eq:Ez}
\rm{CPL:}E^2(z) = & \Omega_m(1+z)^3+(1-\Omega_m) \nonumber\\
           & (1+z)^{3(1+w_0+w_a)}\exp\left[-\frac{3w_az}{1+z}\right]; \nonumber\\
\rm{JBP:}E^2(z) = & \Omega_m(1+z)^3+(1-\Omega_m) \nonumber\\
           & (1+z)^{3(1+w_0)}\exp\left[\frac{3w_az^2}{2(1+z)^2}\right]; \nonumber\\
\rm{Logarithmic:}E^2(z) = & \Omega_m(1+z)^3+(1-\Omega_m)\nonumber\\
                   & (1+z)^{3(1+w_0)}\exp\left[\frac{3w_a}{2}(\ln(1+z))^2\right]; \nonumber\\
\rm{Polynomial:}E^2(z) = & \Omega_m(1+z)^3+(1-\Omega_m)\nonumber\\
                  & \exp\left[3(w_0+w_a)z+1.5w_az^2\right]. \nonumber\\
\end{align}

The confrontation of the above listed expansion rate $E(z)$ with QSO
data would involve distances $D_A(z)=D(z)/(1+z)$ and
$D_L(z)=D(z)(1+z)$, where the comoving distance reads
\begin{equation}\label{eq:D}
D(z) = \left\lbrace \begin{array}{lll}
\frac{c}{H_0\sqrt{|\Omega_{\rm k}|}}\sinh\left[\sqrt{|\Omega_{\rm k}|}\int_{0}^{z}\frac{dz'}{E(z')}\right]~~{\rm for}~~\Omega_{K}>0,\\
\frac{c}{H_0}\int_{0}^{z}\frac{dz'}{E(z')}~~~~~~~~~~~~~~~~~~~~~~~~~~~~~~{\rm for}~~\Omega_{K}=0, \\
\frac{c}{H_0\sqrt{|\Omega_{\rm k}|}}\sin\left[\sqrt{|\Omega_{\rm k}|}\int_{0}^{z}\frac{dz'}{E(z')}\right]~~~~{\rm for}~~\Omega_{K}<0.\\
\end{array} \right.
\end{equation}
The curvature parameter $\Omega_k$ is expressed in terms of $K$ as
$\Omega_k=-c^2K/(a_0H_0)^2$, where $c$ is the speed of light.
Current cosmological observations favor the flat universe
($\Omega_k=0.001\pm0.002$)\cite{Planck2018}; thus, we assumed the
flat universe in this study. Moreover, ref.\cite{Wei2020} used
time-delay lenses and 1598 QSO[XUV] to estimate the curvature and
found $\Omega_k=-0.01^{+0.18}_{-0.17}$, which supports the validity
of our assumption.

\subsection{Model Constraint and Selection}

As mentioned in Section \ref{sec:calibration}, the strategy employed
herein is to fit the free parameters of the four models introduced
above to the QSO[XUV] and QSO[CRS] data. Therefore, we employed the
likelihoods Eq.(\ref{eq:Lik_XUV}) and Eq.(\ref{eq:Lik_CRS}).
However, we fixed the nuisance parameters at the calibrated values
and let the distances depend on cosmic EoS parameters.

To show the constraint ability of quasar samples in comparison and
combination with other standard cosmological probes, we included the
most recent BAO measurements and assume a Gaussian prior for the
Hubble constant of $\rm H_0=67.32\pm4.7\;km\,s^{-1}\,Mpc^{-1}$ from
the GP $H(z)$ reconstruction. The likelihood function for the
uncorrelated BAO measurements can be expressed as
\begin{equation}\label{eq:Lik_BAO_uc}
\ln\mathcal{L}_{\rm{BAO}}({\bf{p}})=-\frac{1}{2}\sum^5_i\frac{[A_{th}({\bf{p}};z_i)-A_{obs}(z_i)]^2}{\sigma_{A_i}^2}
\end{equation}
and for the correlated measurements
\begin{equation}
\ln\mathcal{L}_{\rm{BAO}}({\bf{p}})=-\frac{1}{2}[\vec{A}_{th}({\bf{p}};z_i)-\vec{A}_{obs}(z_i)]^T
C^{-1} [\vec{A}_{th}({\bf{p}};z_i)-\vec{A}_{obs}(z_i)]
\end{equation}
where $C^{?1}$ is the inverse of the covariance matrix
\begin{equation}
C= \left[
  \begin{array}{cccccc}
    624.707 & 23.729   & 325.332 & 8.34963 & 157.386 & 3.57778  \\
    23.729  & 5.60873  & 11.6429 & 2.33996 & 6.39263 & 0.968056 \\
    325.332 & 11.6429  & 905.777 & 29.3392 & 515.271 & 14.1013  \\
    8.34963 & 2.33996  & 29.3392 & 5.42327 & 16.1422 & 2.85334  \\
    157.386 & 6.39263  & 515.271 & 16.1422 & 1375.12 & 40.4327 \\
    3.57778 & 0.968056 & 14.1013 & 2.85334 & 40.4327 & 6.25936 \\
  \end{array}
\right]
\end{equation}
For both uncorrelated and correlated BAO data, the quantities
$A_{obs}(z_i)$, $\sigma_{A_i}$, and $A_{th}(z_i)$ denote the
measurements, observational uncertainties, and theoretical
expressions, respectively. Correlated quantities include the scaled
transverse comoving distance, $D_M(z)(r_{s,fid}/r_s)$, and the
scaled Hubble parameters, $H(z)(r_s/r_{s,fid})$, from the final
galaxy-clustering data of the Baryon Oscillation Spectroscopic
Survey (BOSS). Uncorrelated quantities include the distance ratio,
$r_s/D_V(z)$, $D_V(z)(r_{s,fid}/r_s)$, the Hubble distance,
$(D_H(z))^{0.7}(D_M(z))^{0.3}/r_s$, and the BAO scale along the line
of sight $c/(r_sH(z))$ measurements from the 6dF Galaxy Survey, the
Sloan Digital Sky Survey (SDSS) Data Release 7, the SDSS-IV extended
Baryon Oscillation Spectroscopic Survey (eBOSS) Data Release 14, the
SDSS Data Release 12, and the SDSS-III BOSS Data Release 11,
respectively. $D_V(z)$, $D_H(z)$, $D_M(z)$, and $H(z)$ denote the
spherically averaged angular-diameter distance, Hubble distance,
transverse comoving distance, and Hubble parameters, respectively.
The quantity $r_s$ denotes the size of the sound horizon at the drag
epoch, and $r_{s,fid}$ is the corresponding quantity from the
original measurement. The detailed expressions and quantities can be
found in the text and Table 1 of ref.\cite{Ryan2019}, respectively.

The best-fitted values of cosmological EoS parameters $\bf{p}$ can
be determined by minimizing the chi-square objective function,
$\chi^2({\bf{p}})=-2\ln\mathcal{L}({\bf{p}})$; i.e., by maximizing
the likelihood function. In the case of joint analysis,
log-likelihoods (or equivalently chi-square functions) should be
included (QSO and BAO data are independent).

As already stressed, the log-likelihoods depend on both the EoS
parameters and the calibrated-nuisance parameters $\nu$. Since they
have been fitted by an independent method with some accuracy, they
should be rather marginalized over. Assuming that the prior
distribution of $\nu$, i.e., $P(\nu)$, is Gaussian with the mean
$\overline{\nu}$ and standard deviation $\sigma_{\nu}$:
\begin{equation}
P(\nu)=\frac{1}{\sqrt{2\pi\sigma^2_{\nu}}}e^{-(\nu-\overline{\nu})^2/(2\sigma^2_{\nu})}
\end{equation}
The posterior likelihood function can be obtained by integrating
$\mathcal{L}(\bf{p},\nu)$ over $P(\nu)$ as follows:
\begin{equation}
\mathcal{L}({\bf{p}})=\int^{\infty}_0\mathcal{L}({\bf{p}},\nu)P(\nu)d\nu
\end{equation}
This is how we treated the calibrated nuisance parameters.
More specifically, we assumed the following Gaussian (prior) distributions for the nuisance parameters:
$H_0 = {\cal N}(67.32, 4.7)$, $\gamma={\cal N}(0.634, 0.011)$,
$\beta = {\cal N}(7.75, 0.34)$ and $\delta = {\cal N}(0.232,
0.005)$. Next, we minimized the chi-square posterior (marginalized
over the nuisance parameters)
$\chi_{min}^2({\bf{p}})=-2\ln\mathcal{L}_{max}({\bf{p}})$, which is
equivalent to maximizing the likelihood.
One of our goals is to quantify the degree of support given to
competing models (EoS parametrizations) by the data. While the
standard metric, like the chi-square per degree of freedom ($\chi^2
/ d.o.f.$), describes the goodness of fit, it is insufficient to
compare the models. Therefore, we employed also two specific model
selection techniques \cite{Liddle2007}: the Akaike Information
Criterion (AIC)
\begin{equation}
AIC=\chi^2_{min}+2k
\end{equation}
and the Bayesian Information Criterion (BIC)
\begin{equation}
BIC=\chi^2_{min}+k\ln N
\end{equation}
The AIC value for a single model is useless; what is useful instead
are the differences, $\Delta_i(AIC) = AIC_i - AIC_{min}$, calculated
over the whole set of alternative candidate models (${\cal M}_i;\;i
= 1, ..., M$) where by $AIC_{min}$ we denote as $min\{AIC_i; \;i =
1, ..., M\}$. The relative strength of the evidence for each model
can be calculated as the likelihood of the model given the data
${\cal L}({\cal M}_i|data) \sim exp(-\frac{1}{2}\Delta_i)$. The
relative likelihoods of the models, ${\cal L}({\cal M}_i|data)$,
normalized to unity are termed Akaike weights $w_i$, i.e.,
$w_i=\exp\{-\frac{1}{2}\Delta_i\}/\sum^{M}_{i=1}\exp\{-\frac{1}{2}\Delta_i\}$.
In Bayesian language, an Akaike weight corresponds to the posterior
probability of a model (under the assumption of equal prior
probabilities). The (relative) evidence for the models can also be
determined by the evidence ratios of model pairs as $\frac{w_i}{w_j}
= \frac{{\cal L}({\cal M}_i|data)}{{\cal L}({\cal M}_j|data)}$. The
evidence ratios is substantiated as odds against the given model
with respect to the best one.

\section{Results and Discussion}\label{sec:discussion}

\begin{figure*}
\begin{center}
\centering
\includegraphics[angle=0,width=75mm]{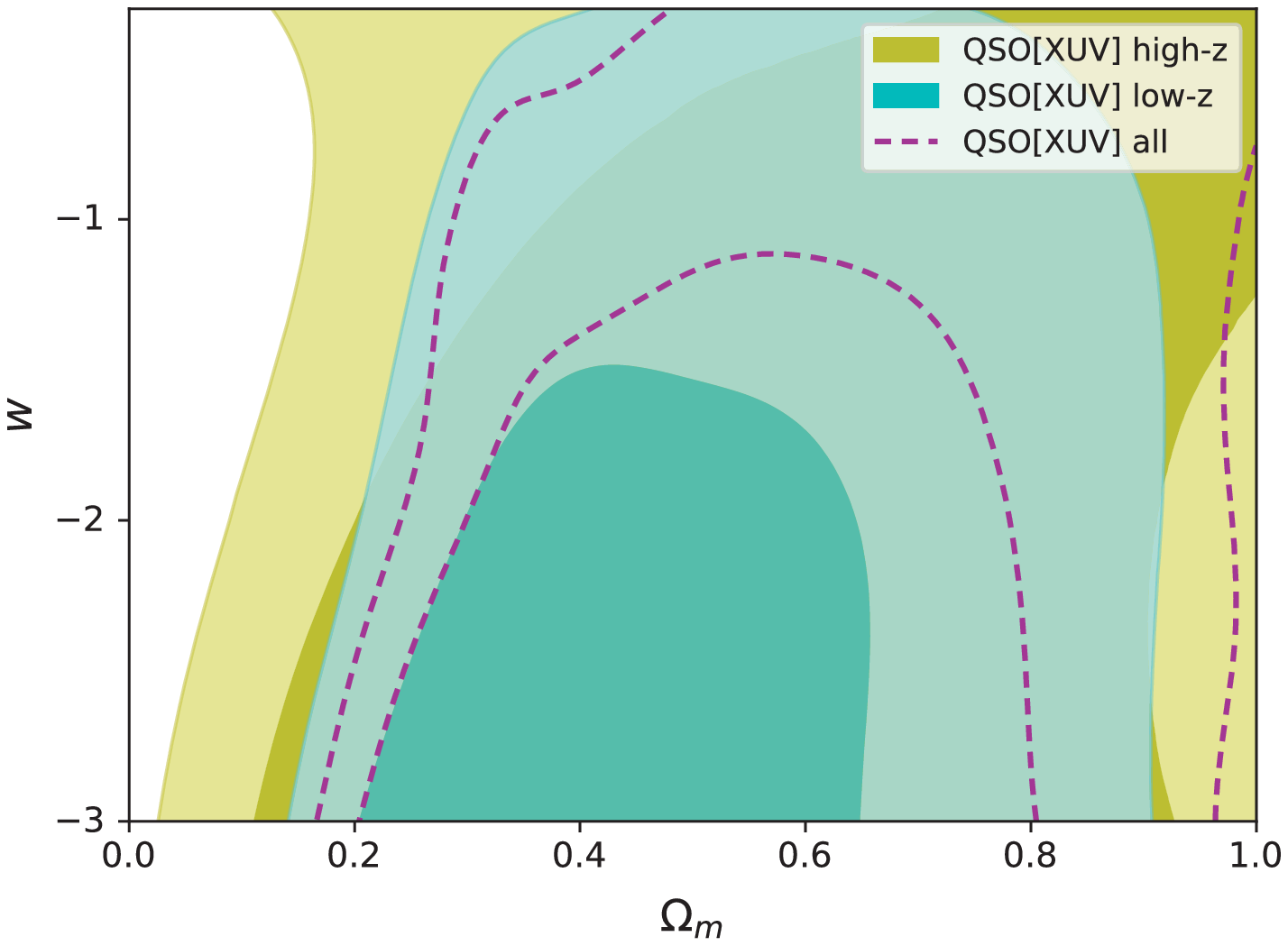}
\includegraphics[angle=0,width=75mm]{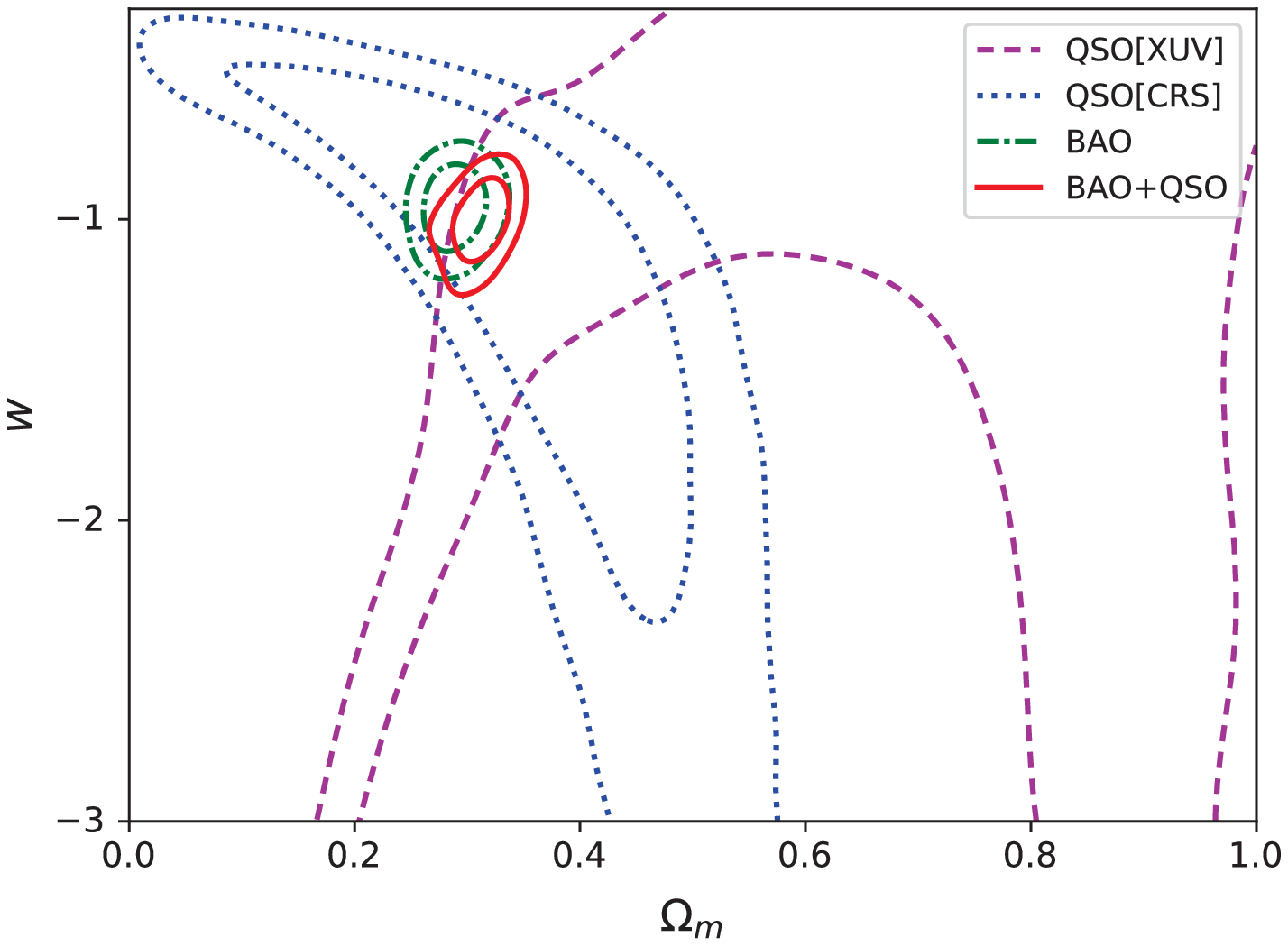}
\caption{\label{fig_wCDM} Two-dimensional distributions of the wCDM
model parameters ($w$ and $\Omega_m$). Left: The cyan and yellow
contours indicate the constraints from 1330 low redshift QSO[XUV]
and 268 high redshift QSO[XUV] data points, respectively. Right: The
magenta dashed, blue dotted, green dash-dot, and red solid contours
indicate the constraints from QSO[XUV], QSO[CRS], BAO, and BAO+QSO,
respectively. Both figures assume the $\rm
H_0=67.32\pm4.7\,km\,s^{-1}\,Mpc^{-1}$ prior derived from Gaussian
Process discussed in Section \ref{sec:calibration}.}
\end{center}
\end{figure*}

\begin{figure}
\begin{center}
\centering
\includegraphics[angle=0,width=75mm]{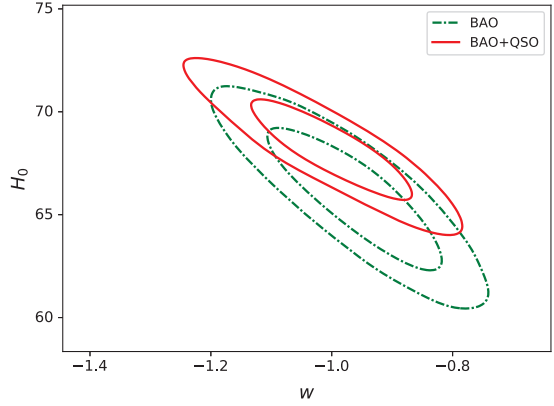}
\caption{\label{fig_H0w} Two-dimensional confidence contours of wCDM
model parameters. The green dash-dot and red solid contours denote
the constraints from BAO and BAO+QSO, respectively.}
\end{center}
\end{figure}

The constraints for the $w$CDM model obtained through distance
measurements of two subsamples of QSO[XUV], independent quasar
samples (XUV or CRS) and both QSO samples combined with BAO are
shown in Fig.(\ref{fig_wCDM}). The 2D marginalized contours with
$1\sigma$ and $2\sigma$ uncertainties from different samples are
shown, both to compare the lower and higher redshift subsamples of
QSO[XUV] and the constraints from different observations. In the
left panel of Fig.(\ref{fig_wCDM}), the filled cyan and the yellow
contours indicate the constraints from the 1330 lower redshift
($z<2$) QSO[XUV] and 268 higher redshifts ($z>2$) QSO[XUV]
subsamples, respectively. Both constraints are consistent with each
other, but the low redshift subsample provided a tighter result.
This implies that extrapolating the calibrated nuisance parameters
from lower to higher redshifts is effective. With the polynomial
expansion $P[\log(1+z)]$, ref.\cite{Risaliti2019} found the
deviation between the concordance $\Lambda$CDM model and the whole
QSO[XUV] data. Our results show that the $\Lambda$CDM model is at
the edge of the $2\sigma$ confidence region, which implies that the
deviation still exists but is more moderate than in
\cite{Risaliti2019}. The validity of the polynomial expansions is
still ambiguous, especially at high redshifts
\cite{Banerjee2020,Yang2020}, and this method requires great care.
In the right panel of Fig.(\ref{fig_wCDM}), the magenta dashed, blue
dotted, green dash-dot, and red solid contours indicate the
constraints from QSO[XUV], QSO[CRS], BAO, and BAO+QSO, respectively.
The corresponding best-fitted values and their $1\sigma$
uncertainties are summarized in Table.\ref{tab_DEEoS}. QSO[XUV]
prefer large matter density and much smaller EoS parameter $w$,
which is consistent with the result reported in
ref.\cite{Khadka2020b}, where the constraining ability of 2015
\cite{Risaliti2015} and 2019 \cite{Risaliti2019} QSO[XUV] samples
were compared. The QSO[CRS] sample showed similar limitations as SN
Ia and preferred too large De EoS parameters. It performed better
than QSO[XUV] as a test for cosmological model parameters. The
results show that adding both QSO[XUV] and QSO[CRS] (hereafter,
abbreviated as QSO) can give much tighter constraint results (see
the comparison of red and green contours in Fig.(\ref{fig_wCDM})).
The degeneracy between the Hubble constant $H_0$ and the DE EoS
parameter $w$ can also be observed in the BAO+QSO sample in
Fig.(\ref{fig_H0w}). To some extent, it alleviates the discrepancy
between the Hubble constant from the cosmological-model-dependent
high-redshift observations and local-model-independent measurements.

It is hard to simultaneously determine the intercept parameter
$\beta$ for QSO[XUV] and the Hubble constant $H_0$ according to
Eq.(\ref{Eq:logFx}). Moreover, the linear-size parameter $l$ for
QSO[CRS] and the Hubble constant $H_0$ are degenerated with each
other according to Eq.(\ref{Eq:thetaz}). Hence, it is difficult to
obtain the $H_0$ value without a good prior knowledge of the
nuisance parameters for QSO[XUV] or QSO[CRS]. Gravitational lensing
provides a promising independent and a rather direct method of
determining the Hubble constant $H_0$, and under the assumption of
different cosmological models, it can be constrained from lensed
quasars if the precision of the measurements is high enough. The
time-delay distance of lensed quasars can be expressed as $D_{\Delta
t} =D_A(z_l)D_A(z_s)/D_A(z_l, z_s)$, where $z_l$ and $z_s$ denote
the redshifts of the lens and the source, respectively. Note that,
in previous studies, another convention with the $1+z_l$ factor
included in the $D_{\Delta t}$ definition was widely used. The
time-delay distance can be derived from time-delay measurements
thus: $\Delta t=(1+z_l)D_{\Delta t}\Delta\phi$, where $\Delta\phi$,
called the Fermat potential difference, depends on the projected
mass distribution of the lens and can be inferred from
high-resolution imaging combined with spectroscopic observations and
stellar kinematics of the lens galaxy. We employed the posterior
probability distribution of $D_{\Delta t}$ and $D_A(z_l)$ from the
H0LiCOW collaboration \cite{Wong2020} and constrain the Hubble
constant under the assumption of the DE EoS parametrizations
discussed herein. The best fitted $H_0$ are $\rm
H^{CPL}_0=81.3^{+5.1}_{-5.4}\;km\,s^{-1}\,Mpc^{-1}$, $\rm
H^{JBP}_0=81.5^{+4.8}_{-5.5}\;km\,s^{-1}\,Mpc^{-1}$, $\rm
H^{Log}_0=81.2^{+5.0}_{-5.0}\;km\,s^{-1}\,Mpc^{-1}$ and $\rm
H^{Poly}_0=99.2^{+26.0}_{-15.3}\;km\,s^{-1}\,Mpc^{-1}$ for the CPL,
JBP, logarithmic, and Polynomial parametrizations, respectively. The
lensed quasar sample can provide a promising constraint for $H_0$,
but there is still a need to improve the precision before it could
be employed in multiparameter cosmological model investigation,
which is predicted to be achieved in the near future. However, in
this study, the Hubble constant priors from the quasar sample were
not employed, and we preferred to adopt $\rm
H_0=67.32\pm4.7\;km\,s^{-1}\,Mpc^{-1}$ from the reconstruction of
the Hubble parameter measurements. The BAO itself also favored
smaller Hubble constants, which is consistent with previous study
\cite{Khadka2020b,Zhang2020}.

\begin{figure*}
\begin{center}
\centering
\includegraphics[angle=0,width=75mm]{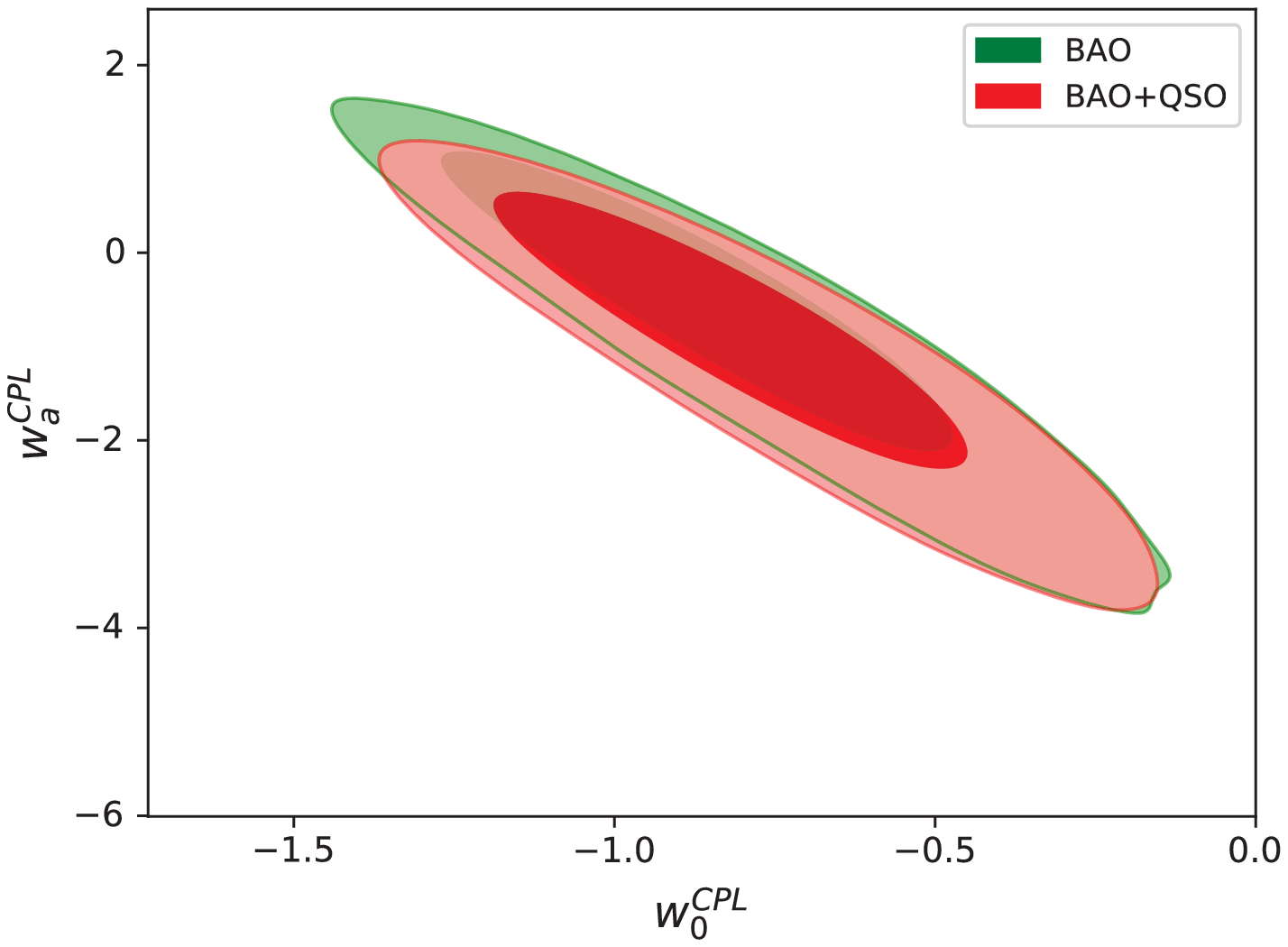}
\includegraphics[angle=0,width=75mm]{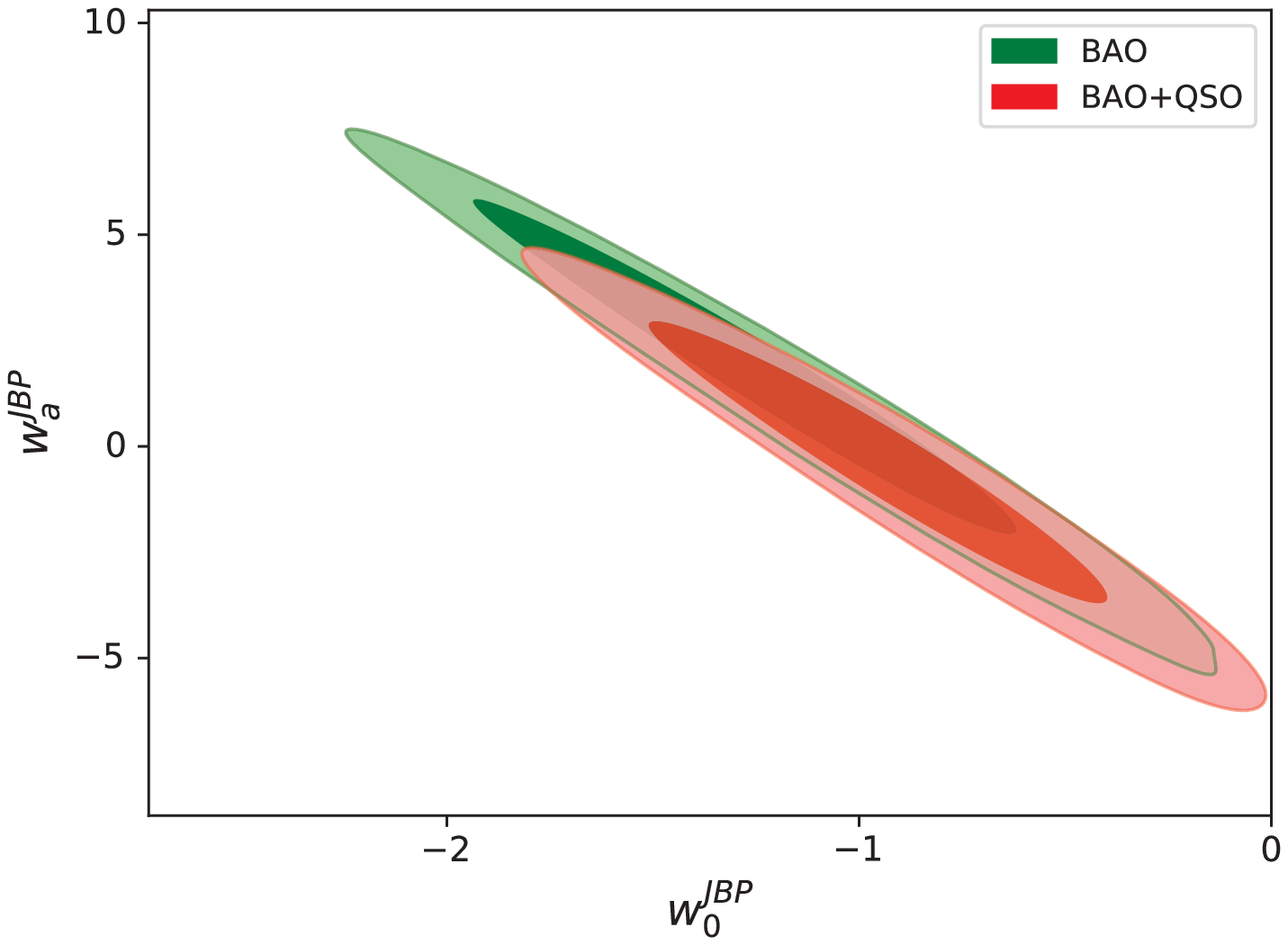}
\includegraphics[angle=0,width=75mm]{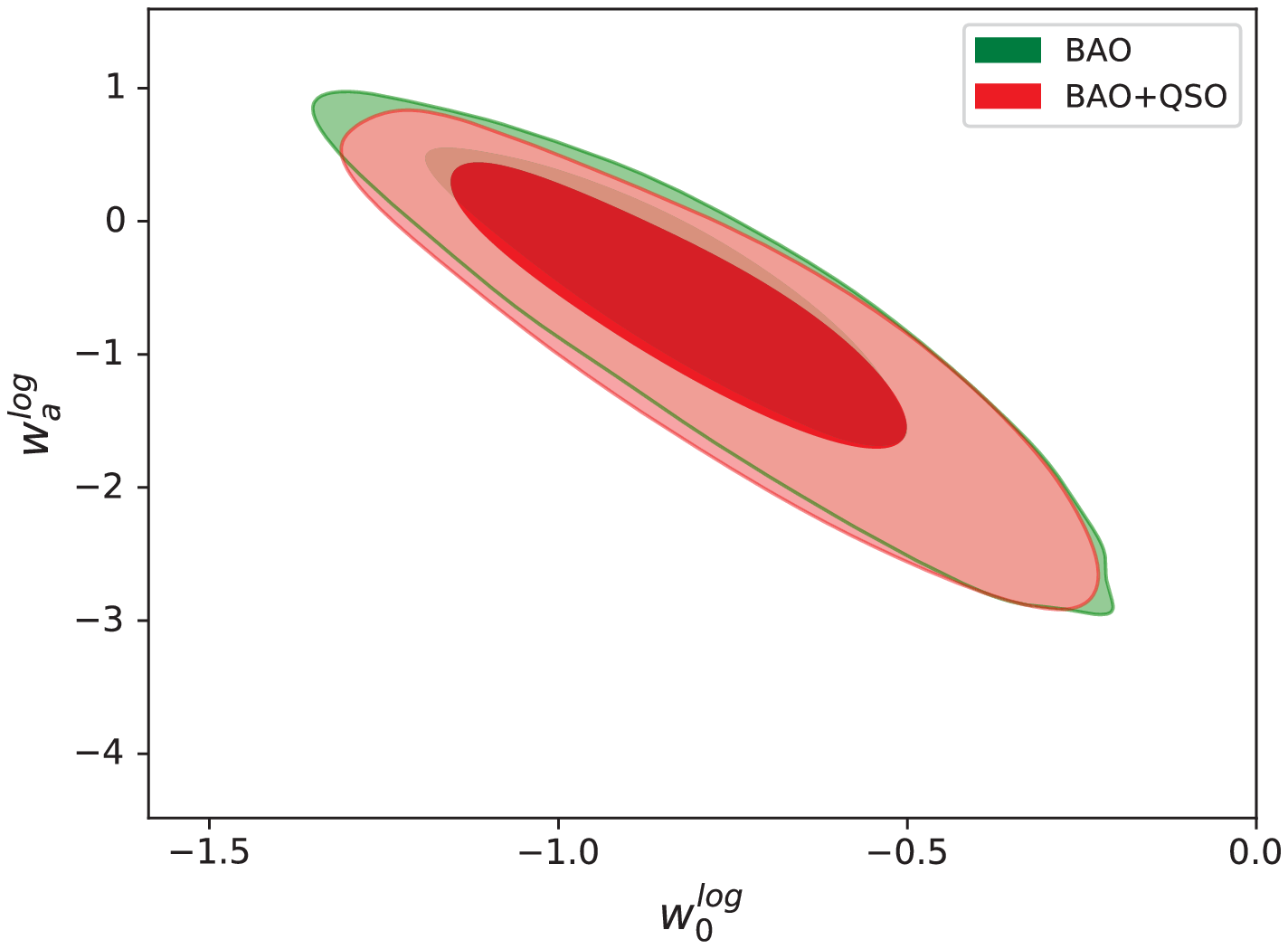}
\includegraphics[angle=0,width=75mm]{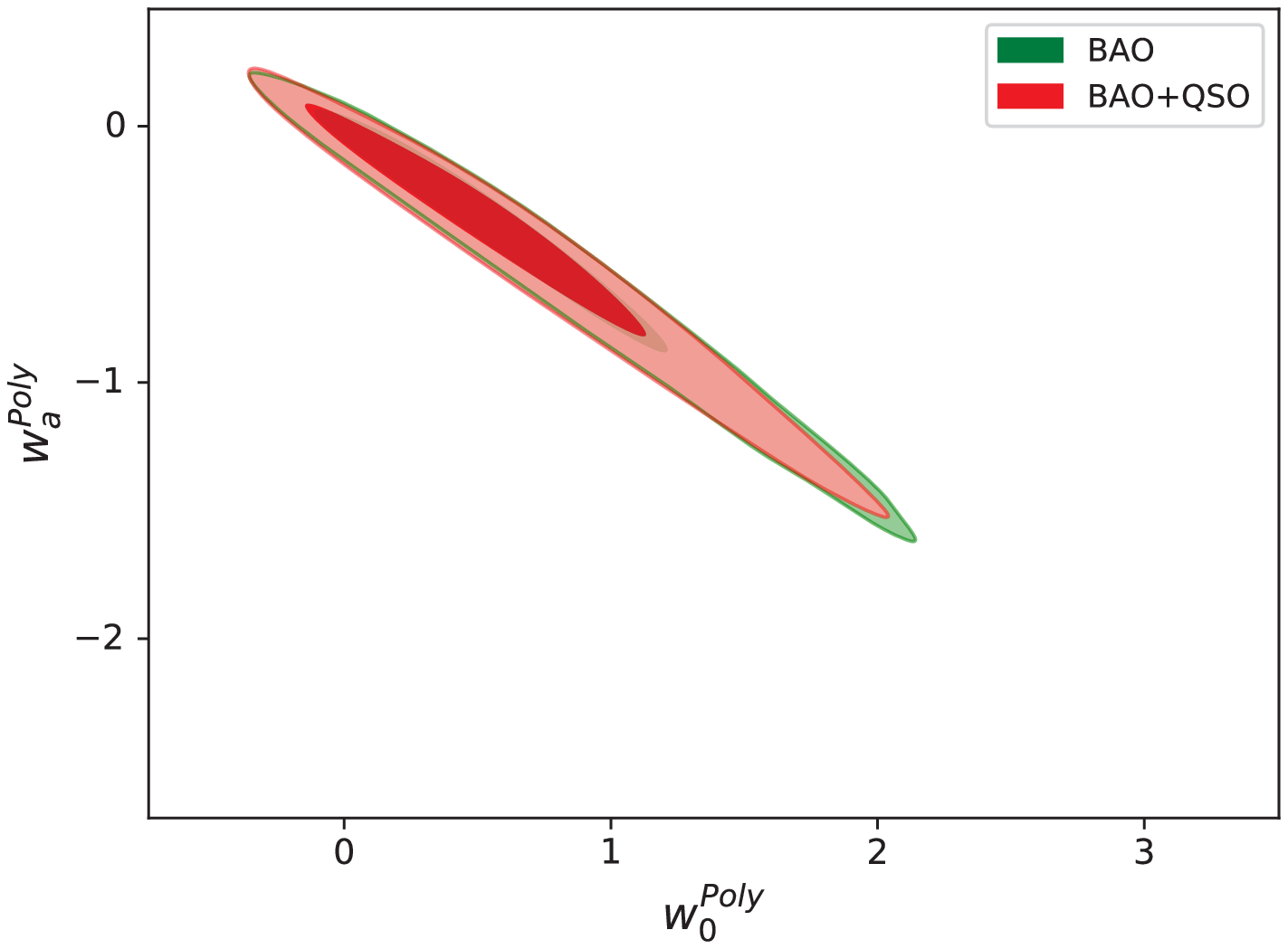}
\caption{\label{fig_w0wa} Dark Energy Equation of State parameters
constraint results with $\rm H_0=67.32\pm4.7\,km\,s^{-1}\,Mpc^{-1}$
prior. The green and red contours denote BAO and BAO+QSO,
respectively. }
\end{center}
\end{figure*}

\begin{figure}
\begin{center}
\centering
\includegraphics[angle=0,width=90mm]{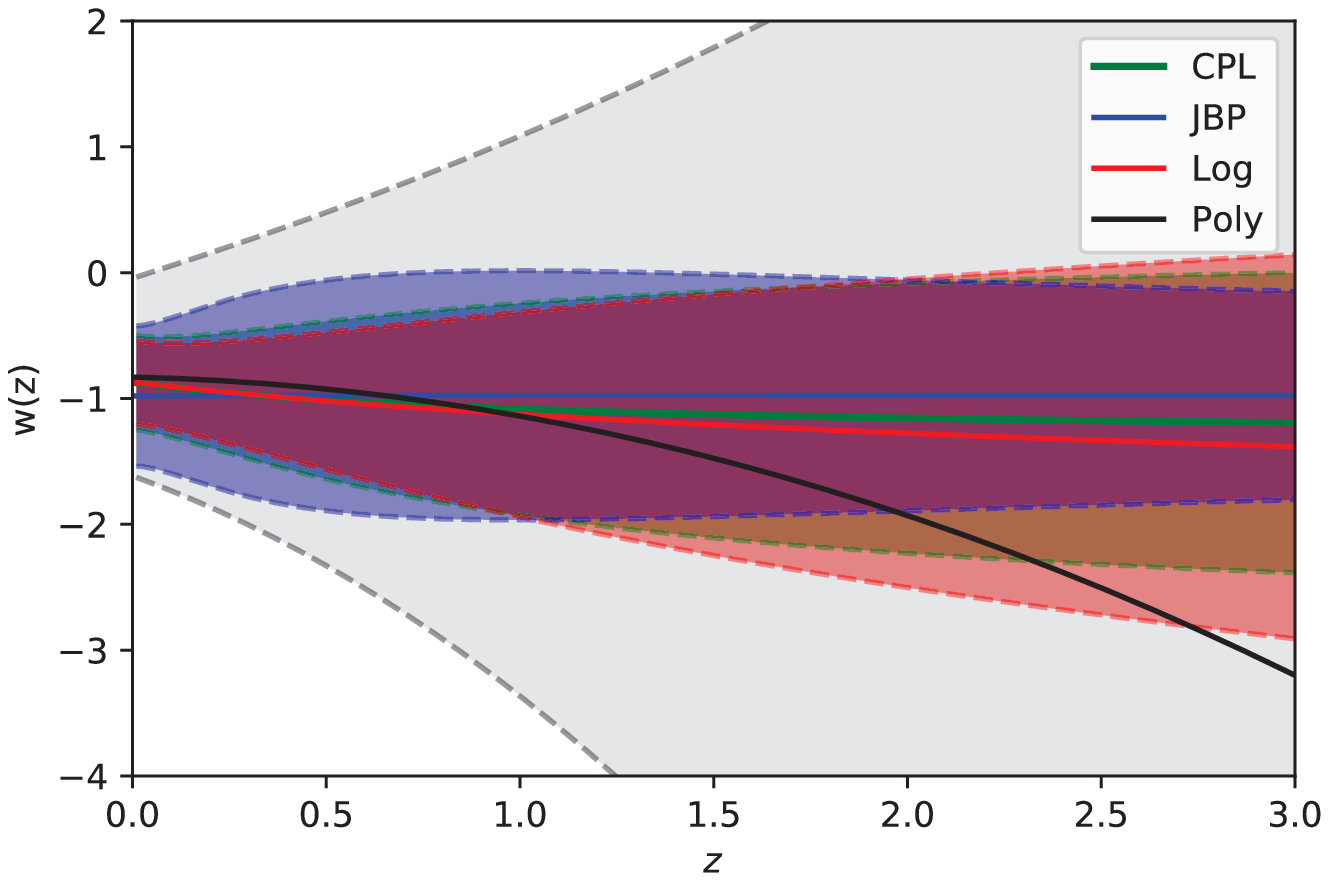}
\caption{\label{fig_wz} Evolution of Dark Energy Equation of State
parameters according to different parameterizations following the
joint BAO+QSO constraint.}
\end{center}
\end{figure}

Our main focus was to investigate the DE EoS parameters; therefore,
we used the Hubble parameters and BAO measurements to get tighter
constraints for the Hubble constant $H_0$ and the matter density
$\Omega_m$. Obviously, the QSO sample can improve the constraining
accuracy for all the parametrizations discussed in this work, as
observed in Fig.(\ref{fig_w0wa}). For comparison, we also included
the $\Lambda$CDM model in this study. The detailed results are
summarized in Table.(\ref{tab_DEEoS}). For more understanding,
Fig.(\ref{fig_wz}) shows the evolution of the DE EoS parameters
according to different parameterizations following the joint BAO+QSO
constraints. From the AIC and BIC criteria for the BAO+QSO
constraints, the Polynomial parametrization is most favored, whereas
the JBP parametrization is least favored among the two-parameter DE
EoS parametrizations. The standard $\Lambda$CDM model exhibited the
best performance. To quantify the constraining power of the quasar
sample, we calculated the figure-of-merit (FoM) for the DE models
with BAO only and BAO+QSO combined. The FoM can be expressed as
\begin{equation}
FoM=(\det Cov(\textbf{p}))^{-1/2}
\end{equation}
where $Cov(\textbf{p})$ is the covariance matrix of relevant
cosmological parameters $\textbf{p}$. The final FoM results are
listed in Table(\ref{tab_FoM}), which also lists the relevant
results of the information criteria and weights.

\begin{table}
\begin{center}
\caption{Best-fitted parameters for different dark energy models. We
assumed the Gaussian prior distributions for the nuisance
parameters: $H_0 = {\cal N}(67.32, 4.7)$, $\gamma={\cal N}(0.634,
0.011)$, $\beta = {\cal N}(7.75, 0.34)$ and $\delta = {\cal
N}(0.232, 0.005)$.}
{{\scriptsize%\footnotesize\tiny
 \begin{tabular}{c l c c c c c c c c c c c} \hline\hline
Model & Data & $\Omega_m$ & $w_0$ & $w_a$ & $\chi^2/d.o.f$  \\
\cline{1-6}
$\Lambda$CDM & BAO & $0.28^{+0.03}_{-0.03}$  & $-1$                     & $-$      & $7.28/10$   \\
wCDM         & BAO & $0.29^{+0.03}_{-0.03}$  & $-0.95^{+0.14}_{-0.15}$  & $-$      & $7.32/9$    \\
CPL  & BAO & $0.30^{+0.06}_{-0.06}$ & $-0.88^{+0.42}_{-0.37}$ & $-0.27^{+1.29}_{-1.98}$  & $7.31/8$ \\
JBP  & BAO & $0.27^{+0.06}_{-0.07}$ & $-1.23^{+0.63}_{-0.69}$ & $1.99^{+3.44}_{-4.39}$   & $6.39/8$ \\
Log  & BAO & $0.31^{+0.05}_{-0.05}$ & $-0.84^{+0.35}_{-0.34}$ & $-0.29^{+0.79}_{-1.54}$  & $7.18/8$ \\
Poly & BAO & $0.32^{+0.04}_{-0.05}$ & $0.38^{+0.86}_{-0.49}$  & $-0.21^{+0.28}_{-0.69}$  & $6.31/8$ \\
\hline
$\Lambda$CDM & BAO+QSO        & $0.31^{+0.03}_{-0.03}$  & $-1$ & $-$            & $1967.05/1728$                    \\
wCDM         & BAO+QSO        & $0.31^{+0.03}_{-0.03}$  & $-1.00^{+0.14}_{-0.13}$  & $-$            & $1967.03/1727$                      \\
CPL  & BAO+QSO & $0.33^{+0.04}_{-0.04}$ & $-0.87^{+0.44}_{-0.31}$ & $-0.43^{+1.14}_{-1.88}$  & $1966.65/1726$   \\
JBP  & BAO+QSO & $0.31^{+0.05}_{-0.04}$ & $-0.98^{+0.61}_{-0.50}$ & $0.02^{+2.87}_{-3.69}$  & $1967.14/1726$  \\
Log  & BAO+QSO & $0.33^{+0.03}_{-0.03}$      & $-0.87^{+0.38}_{-0.28}$ & $-0.37^{+0.79}_{-1.36}$ & $1966.33/1726$   \\
Poly & BAO+QSO & $0.33^{+0.04}_{-0.04}$     & $0.41^{+0.74}_{-0.54}$  & $-0.24^{+0.32}_{-0.59}$ & $1964.99/1726$   \\
\hline \hline
\end{tabular}} \label{tab_DEEoS}}
\end{center}
\end{table}

\begin{table*}
\begin{center}
\caption{AIC and BIC information criteria (values, differences,
weights) and the Figure-of-Merit for DE EoS parameters (pairwise and
for all parameters).} {{\scriptsize
 \begin{tabular}{c l c c c c c c c c c} \hline\hline
Model & Data & $\rm AIC_i$ & $\Delta_i$(AIC) & $w_i$(AIC) & $\rm
BIC_i$ & $\Delta_i$(BIC) & $w_i$BIC & FoM($w_0$-$w_a$) &
FoM($\bf{p}$) \\ \cline{1-10}
$\Lambda$CDM & BAO & $9.28$   & $0$    & $0.4838$ & $9.68$  & $0$    & $0.5646$  & $-$    & $51.9$ \\
wCDM         & BAO & $11.32$  & $2.04$ & $0.1744$ & $12.12$ & $2.44$ & $0.1667$  & $-$    & $601.0$ \\
CPL          & BAO & $13.31$  & $4.03$ & $0.0645$ & $14.50$ & $4.82$ & $0.0507$  & $8.4$  & $439.1$  \\
JBP          & BAO & $12.39$  & $3.11$ & $0.1022$ & $13.58$ & $3.90$ & $0.0803$  & $4.4$  & $338.0$  \\
Log          & BAO & $13.18$  & $3.90$ & $0.0688$ & $14.37$ & $4.69$ & $0.0541$  & $11.6$ & $592.7$  \\
Poly         & BAO & $12.31$  & $3.03$ & $0.1063$ & $13.50$ & $3.82$ & $0.0836$  & $29.5$ & $2046.0$ \\
\hline
$\Lambda$CDM & BAO+QSO & $1969.05$ & $0$    & $0.4466$ & $1974.51$ & $0$     & $0.9727$ & $-$    & $95.1$  \\
wCDM         & BAO+QSO & $1971.03$ & $1.98$ & $0.1659$ & $1981.94$ & $7.43$  & $0.0237$ & $-$    & $941.2$ \\
CPL          & BAO+QSO & $1972.65$ & $3.60$ & $0.0738$ & $1989.02$ & $14.51$ & $0.0007$ & $8.9$  & $827.0$  \\
JBP          & BAO+QSO & $1973.14$ & $4.09$ & $0.0578$ & $1989.51$ & $15.00$ & $0.0005$ & $4.5$  & $418.0$  \\
Log          & BAO+QSO & $1972.33$ & $3.28$ & $0.0866$ & $1988.70$ & $14.19$ & $0.0008$ & $12.0$ & $1101.1$ \\
Poly         & BAO+QSO & $1970.99$ & $1.94$ & $0.1693$ & $1987.36$ & $12.85$ & $0.0016$ & $30.5$ & $3049.9$ \\
\hline \hline
\end{tabular}} \label{tab_FoM}}
\end{center}
\end{table*}

\section{Conclusion}\label{sec:conclusion}

Quasars are one of the brightest objects in the universe, and their
cosmological redshift can be larger than $z=7$. Therefore, they are
natural candidates for standard cosmological probes. Several
attempts have been made to use quasars as cosmological probes;
however, their practical use as standard probes is still
challenging. Currently, two classes of quasar samples appear to be
promising. First, the nonlinear relation between the ultraviolet and
X-ray luminosities of quasars (QSO[XUV] in short) can be used to
derive luminosity distances. Second, the linear size of the compact
radio structure in quasars (QSO[CRS] in short) can provide
angular-diameter distances. These two processes have their
respective advantages; however, they are challenged by their
nuisance parameters, which demand clarification before they can be
employed in any cosmological applications. In this study, we
simultaneously refreshed the calibration of such parameters through
a cosmological-model independent method in light of the newly
compiled Hubble parameter sample for the two processes. The
calibration results for the QSO[XUV] nuisance parameters are: the
slope parameter $\gamma=0.634\pm0.011$, the intercept
$\beta=7.75\pm0.34$, and the dispersion $\delta=0.232\pm0.005$.
These results are consistent with those of ref.\cite{Risaliti2019}
and fitted with the cosmological parameters with the nuisance
parameters simultaneously \cite{Khadka2020b}. For the QSO[CRS]
linear-size Parameter, $l=10.89\pm0.19\;pc$, which is a little
smaller than that calibrated with Hubble parameters in a redshift
coverage of $z<1.2$ \cite{Cao2017a} and Union2.1 SN Ia in a redshift
coverage of $z<1.4$ \cite{Cao2017b}.

With the calibrated nuisance parameters, we tested and compared
their abilities to constrain cosmological models. The results show
that both quasar samples are promising complementary probes, and
with the current measurement precision, the compact radio quasars
perform better. The considerable degeneracy between $w$ and $H_0$
shows the importance of using complementary $H_0$ measurements from
quasar observations, such as gravitational lensing time-delay
measurements, before DE EoS investigations. Both Hubble parameter
measurements and BAO observations prefer smaller Hubble constants,
and thus, we considered the $H_0$ derived from the reconstruction of
the Hubble parameter as a prior during the investigation. Moreover,
with the calibrated quasar sample, we investigated four different DE
EoS with $H_0$ derived from the Gaussian Process reconstruction, the
most recent baryon acoustic oscillation measurements, and both
quasar samples. The results show that the quasar sample can improve
the precision and the combined QSO+BAO measurements are consistent
with the standard $\Lambda$CDM model. From the AIC and BIC values,
among the four parametrizations discussed, the polynomial
parametrization is most favored, whereas the JBP parametrization is
least favored. In the future, samples of compact radio quasars and
X-ray UV quasar data that are better controlled for systematics and
with significantly lower dispersion would be helpful in the
high-precision studies of dynamical DE models.

\acknowledgements{This work was supported by the National Natural
Science Foundation of China under Grant Nos. 12021003, 11690023,
11633001 and 11920101003, the National Key R\&D Program of China
(Grant No. 2017YFA0402600), the Beijing Talents Fund of Organization
Department of Beijing Municipal Committee of the CPC, the Strategic
Priority Research Program of the Chinese Academy of Sciences (Grant
No. XDB23000000), the Interdiscipline Research Funds of Beijing
Normal University, and the Opening Project of Key Laboratory of
Computational Astrophysics, National Astronomical Observatories,
Chinese Academy of Sciences. M.B. was supported by the Foreign
Talent Introducing Project and Special Fund Support of Foreign
Knowledge Introducing Project in China. He was supported by the Key
Foreign Expert Program for the Central Universities No. X2018002. X.
Li was supported by the National Natural Science Foundation of China
(Grant Nos. 11947091, 12003006) and Hebei NSF (Grant No.
A202005002). }

\end{document}